\documentclass[a4paper,fleqn]{cas-sc}

\usepackage[numbers,sort&compress]{natbib}
\usepackage{graphicx}
\usepackage{caption}
\usepackage{subcaption}
\usepackage{placeins}
\usepackage{float}
\usepackage{needspace}

\def\tsc#1{\csdef{#1}{\textsc{\lowercase{#1}}\xspace}}
\tsc{WGM}
\tsc{QE}
\tsc{EP}
\tsc{PMS}
\tsc{BEC}
\tsc{DE}

\begin{document}
\let\WriteBookmarks\relax
\def\floatpagepagefraction{1}
\def\textpagefraction{.001}
\shorttitle{Interpretable bearing remaining useful life prediction method}
\shortauthors{Authors}

\title [mode = title]{A physics-enhanced bidirectional multi-order graph fusion network for interpretable bearing remaining useful life prediction}                      

\author[1]{Haoxuan Zhang}

\author[1]{Dinghao Yang}

\author[1]{Kangning Zhang}

\author[2]{Shaoyong Guo}

\author[3]{Haisheng Li}

\author[1]{Rui Yang}[orcid=0009-0009-5211-242X]
\cormark[1]

\author[1]{Ruijun Liu}

\affiliation[1]{organization={School of Software, Beihang University},
                city={Beijing},
                postcode={100191}, 
                country={China}}

\affiliation[2]{organization={State key Laboratory of Networking and Switching Technology, Beijing University of Posts and Telecommunications},
                city={Beijing},
                postcode={100876}, 
                country={China}}

\affiliation[3]{organization={School of Computer and Artificial Intelligence, Beijing Technology and Business University},
                city={Beijing},
                postcode={100048}, 
                country={China}}

\cortext[cor1]{Corresponding author, email: yang.rui@buaa.edu.cn}

\begin{abstract}
Accurate prediction of bearing remaining useful life (RUL) is a key challenge for intelligent maintenance. Although deep learning-based prediction methods have showed effectiveness, existing methods still have limitations in learning nonlinear bearing degradation processes and model interpretability. Especially in engineering applications, the "black box" nature of deep learning models can easily raise concerns about their reliability. Therefore, we propose a physics-enhanced bidirectional multi-order graph fusion network for interpretable bearing RUL prediction. Our network mines complementary information from both forward and backward degradation sequences. Specifically, our network introduces a multi-order graph propagator to capture the local-global degradation dependencies. A gated cross-fusion mechanism is further designed to dynamically balance the feature contributions from both forward and backward directions. Then, our network stores representative historical degradation prototypes in dynamic memory, so that the final RUL prediction no longer depends solely on the current latent features, but is guided by reusable historical degradation knowledge. To reveal how our model learns the nonlinear degradation process, the feature mapping parts utilize the Kolmogorov-Arnold network, which allows the nonlinear mapping to be visualized using learnable functions. Finally, a physics-enhanced dynamic loss function is developed to help our network learn effective and reliable degradation representations. Extensive experiments on two public datasets show that our method achieves the lowest error while providing more conservative estimates than existing methods. Our code is available at https://github.com/IMGresearcher/PE-BMGN.
\end{abstract}


\maketitle

\section{Introduction}

With the rapid development of intelligent manufacturing and industrial equipment digitalization, prognostics and health management (PHM) has become an essential technology for improving the reliability, safety, and maintainability of modern mechanical systems \cite{yucesan2021survey,chia2024review,zhang2024hybrid,salinas2025comprehensive}. As one of the core tasks in PHM, remaining useful life (RUL) prediction aims to estimate the future degradation trajectory and failure time of key components based on condition monitoring data, thereby providing decision support for predictive maintenance. Accurate RUL prediction can reduce unexpected downtime, avoid catastrophic failures, and optimize maintenance schedules by shifting maintenance strategies from passive repair to proactive health management \cite{lin2024advancing,gao2025vsc,kim2025fisher,zhang2026discovery}. Therefore, developing interpretable RUL prediction models has become an important research direction in the lifecycle management of high-value industrial equipment \cite{zhao2026low,liu2026remaining}.

Rolling bearings are critical components widely used in rotating machinery, where they undertake load transmission and motion support. Their degradation state directly affects the operational stability of the whole mechanical system \cite{rehman2025deep,wang2025intelligent,bai2025spatio,bao2026theory}. However, bearing degradation is usually affected by complex working conditions, alternating loads, material fatigue, lubrication variation, and environmental disturbances, resulting in strong nonlinearity, non-stationarity, and individual variability in vibration signals. In practical scenarios, early degradation signatures are often weak and easily submerged by noise, while middle- and late-stage degradation may exhibit abrupt changes and stage-dependent correlations. These characteristics make it difficult to establish a robust mapping from raw monitoring signals to RUL values. In the context of smart manufacturing and Industry 5.0 \cite{wang2024automatic,puthanveettil2025review,jing2026human}, the key challenge of bearing RUL prediction lies in how to extract degradation-sensitive representations from high-dimensional non-stationary signals and how to model the evolving nonlinear dependencies across different degradation stages in an interpretable manner.

Existing RUL prediction studies can be roughly divided into two categories. The first category is data-driven methods, which directly learn the mapping from monitoring signals or extracted features to RUL values through neural networks. Yang et al. \cite{yang2022bearing} proposed a regression shapelet-based graph neural network method, referred to as T-GCN in the comparative experiments, which constructs graph structures from regression shapelets and combines graph convolution with recurrent units to model spatio-temporal degradation information. Yang et al. \cite{yang2022node} proposed ChebGCN-LSTM, where node-level path graphs are constructed to represent chronological relationships among discrete signals, and Chebyshev graph convolution is combined with BiLSTM for bearing RUL prediction. Wei et al. \cite{wei2023bearing} proposed SAGCN-SA, a self-adaptive graph convolutional network with a self-attention mechanism, to capture correlations among features at different time points without relying on recurrent architectures. Zhao et al. \cite{zhao2025node} proposed STACP-GCN, which further improves graph-based degradation modeling by combining time-frequency feature extraction, adaptive graph construction, and clustering graph pooling. These studies demonstrate that CNNs \cite{gu2018recent}, RNNs \cite{salehinejad2017recent}, LSTMs \cite{van2020review}, attention mechanisms \cite{niu2021review}, and GCNs \cite{zhang2019graph} can effectively enhance degradation representation learning. However, despite differences in feature extraction structures, the final degradation-to-RUL mapping in these models is still mainly realized by implicit MLP-like fully connected transformations. As a result, the learned nonlinear mapping lacks explicit functional form, making it difficult to explain how specific degradation features contribute to the final RUL estimation.

The second category is physics-guided or physics-informed methods, which incorporate signal processing priors, physical degradation information, or physical constraints into data-driven frameworks. Xu et al. \cite{xu2023multi} proposed MR-LSTM, which uses discrete wavelet transform to generate multi-resolution sequences and then employs a two-layer LSTM to learn temporal dependencies across different resolutions.  Yang et al. \cite{yang2023dual} proposed SDFEPN, which introduces Fourier and wavelet enhancement blocks to capture global degradation trends and local details, thereby embedding frequency-domain physical rules into the neural prediction framework. He et al. \cite{he2025prediction} proposed DC-DGCN, which constructs a two-stage updated digital twin model to describe bearing defect evolution and integrates the mapped physical defect information into graph-based RUL prediction. More recently, Zhao et al. \cite{zhao2025new} proposed DyWave-BiAGCN, which introduces dynamically differentiable wavelet decomposition and physical information constraints to improve the adaptability and physical consistency of bearing RUL prediction. These methods improve the physical plausibility of feature extraction and provide a certain degree of interpretability through decomposition, digital twin modeling, or physical regularization. Nevertheless, their prediction backbones still largely depend on conventional neural transformations, especially MLP-like regression heads, to approximate the relationship between degradation representations and RUL values. Thus, the interpretability improvement mainly lies in the training constraint level, while the nonlinear decision process inside the prediction model remains insufficiently transparent.

\begin{figure}
	\centering
	\includegraphics[width=.98\textwidth]{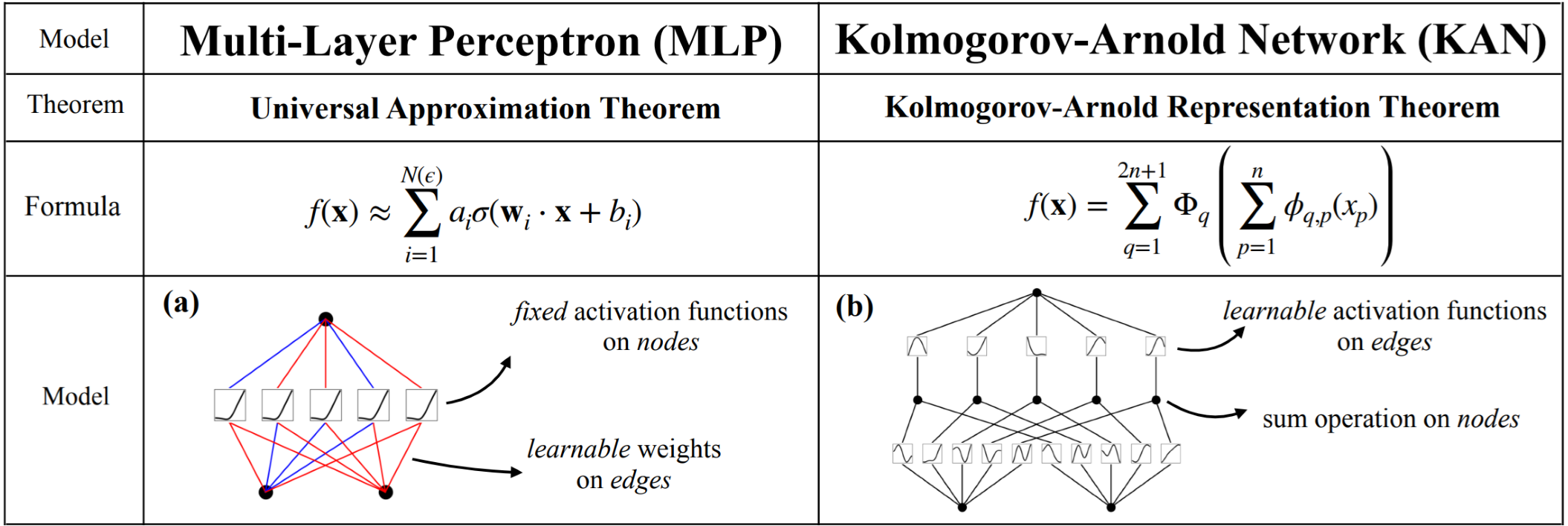}
	\caption{Architectural comparison of Multi-Layer Perceptrons and Kolmogorov-Arnold Networks. As can be seen from the visualization of the model, KAN transforms the basic modeling unit of the model from scalar weights that cannot be directly interpreted into a visualized and analyzable one-dimensional function mapping.}
	\label{FIG:0}
\end{figure}

Liu et al. \cite{liu2025kan} proposed Kolmogorov-Arnold Network (KAN), which provides a promising alternative to the conventional MLP for nonlinear degradation-to-RUL mapping. As shown in Figure~\ref{FIG:0}, MLPs are generally constructed by stacking linear transformations and fixed nonlinear activation functions, where the nonlinear mapping ability is mainly produced by node-wise activations and dense weight matrices. Although this structure has strong approximation capability, the learned decision process is usually distributed across numerous implicit weights. In contrast, KAN is inspired by the Kolmogorov-Arnold representation theorem and replaces scalar linear weights with learnable univariate functions. In a KAN layer, each edge corresponds to a trainable nonlinear function, while each node mainly performs summation over incoming transformed signals. Therefore, KAN changes the basic modeling unit from an implicit weight coefficient to an explicit functional mapping. In KAN, the one-dimensional function corresponding to each edge can be directly visualized, allowing observation of its monotonicity, local peaks, saturation intervals, inflection points, and nonlinear curvature, thereby analyzing the specific impact of a given input feature on the output. In contrast, in MLP, individual weights typically cannot independently correspond to explicit meanings, and their decision-making process is more akin to a black box mapping.

Several studies have introduced KAN into RUL prediction tasks. Zheng et al. \cite{ZHENG2026337} proposed TC-BiKAN for online cross-condition bearing RUL prediction. In this framework, KAN is mainly used in the nonlinear regression stage: a BiLSTM is first adopted to extract bidirectional temporal degradation features, and the conventional fully connected regression head is then replaced by KAN-based learnable spline functions. Shen et al. \cite{shen2025sparse} proposed SSPGKAN, a siamese KAN is developed to classify bearing degradation stages, and independent stage path graphs are then constructed for different degradation stages. Zhou et al. \cite{zhou2025aero} proposed a DCRA-KAN model for aero-engine RUL prediction, where deformable convolutional residual attention modules are used to extract multi-scale degradation features, and KAN serves as the final regressor to improve the mapping from extracted features to RUL values.

Existing KAN-based RUL methods primarily incorporate KAN directly into the representation module or the final prediction module, without considering how to design a module better suited to KAN to enhance the overall framework's ability to model nonlinear degradation processes. Meanwhile, KAN's greatest advantage lies in placing learnable activation functions on the edges, improving model interpretability. However, these methods have not addressed the interpretability of the proposed model.

To address these issues, we propose PE-BMGN, an interpretable bearing RUL prediction method based on physics-enhanced bidirectional multi-order graph fusion network. PE-BMGN does not simply integrate KAN into the existing RUL prediction method. Instead, based on the characteristics of bearing life cycle evolution, it embeds KAN into the mapping process from graph features to degradation features and the mapping process from degradation features to RUL. By leveraging its learnable activation function, PE-BMGN is able to represent nonlinear degradation mappings at different stages in a more explicit and analytical form. This design not only enhances the model's ability to capture complex degradation pattern, but also provides more transparent prediction evidence for engineering applications. Our main contributions are:

\begin{itemize} 
\item We propose a bidirectional multi-order graph fusion network for degradation dependency modeling. The degradation graph is decomposed into forward and backward propagation views, where Chebyshev graph convolution is utilized to capture multi-order local-to-global degradation correlations. And we design a gated cross-fusion mechanism to adaptively balance forward and backward information.

\item We introduce a memory-augmented prediction module for reliable RUL estimation. Representative historical degradation prototypes are stored in a dynamic memory bank, and reusable degradation knowledge is retrieved according to the current degradation state. In this way, the final prediction is further guided by historically similar degradation knowledge.

\item We develop a physics-enhanced dynamic loss function to guide effective and reliable representation learning. Building upon the physical constraints, we further constrain the Kolmogorov-Arnold spline coefficients, enabling the network to generate effective and reliable functional responses.

\item A comprehensive evaluation of the proposed PE-BMGN based on XJTU-SY and PHM2012 bearing datasets, demonstrating superior performance across all operating conditions. Interpretability experiments reveal how our model learns the degradation process, providing a new perspective on RUL prediction.

\end{itemize}

The rest of this paper is organized as follows. Section~2 formulates the bearing RUL prediction problem. Section~3 presents the proposed PE-BMGN framework in detail. Section~4 conducts comprehensive experiments on the XJTU-SY and PHM2012 bearing datasets. Section~5 summarizes the conclusions of this paper and discusses potential future research directions.

\section{Problem Definition}

The RUL prediction task can be formulated as learning a mapping from the input sequence to the predicted normalized RUL sequence. For RUL prediction, a complete degradation run of one bearing is regarded as a temporal sample. Let the training set be denoted by
\begin{equation}
\mathcal{D}=\left\{\left(\mathcal{X}^{(b)}, \mathbf{y}^{(b)}\right)\right\}_{b=1}^{N},
\end{equation}
where $N$ is the number of bearing runs. For the $b$-th run, the input sequence is written as $\mathcal{X}^{(b)}=[\mathbf{s}^{(b)}_{1},\mathbf{s}^{(b)}_{2},\ldots,\mathbf{s}^{(b)}_{T_b}] \in \mathbb{R}^{T_b \times C \times L}$, where $T_b$ is the number of monitoring steps, $C$ is the number of signal channels, and $L$ is the signal length of each acquisition. In this paper, $C=2$, corresponding to the horizontal and vertical vibration signals. The corresponding target sequence is denoted by $\mathbf{y}^{(b)}=[y^{(b)}_{1},y^{(b)}_{2},\ldots,y^{(b)}_{T_b}]^\top \in \mathbb{R}^{T_b}$. To facilitate unified regression across different bearings, the RUL labels are normalized into the interval $[0,1]$, where the label at the $t$-th monitoring step is defined as $y^{(b)}_{t}=(T_b-t)/(T_b-1)$. Accordingly, the first monitoring step corresponds to 1 and the last monitoring step corresponds to 0. For clarity of notation, the azimuth indicator $(b)$ is omitted in the subsequent section.

\section{Methodology}

\begin{figure}
	\centering
	\includegraphics[width=.98\textwidth]{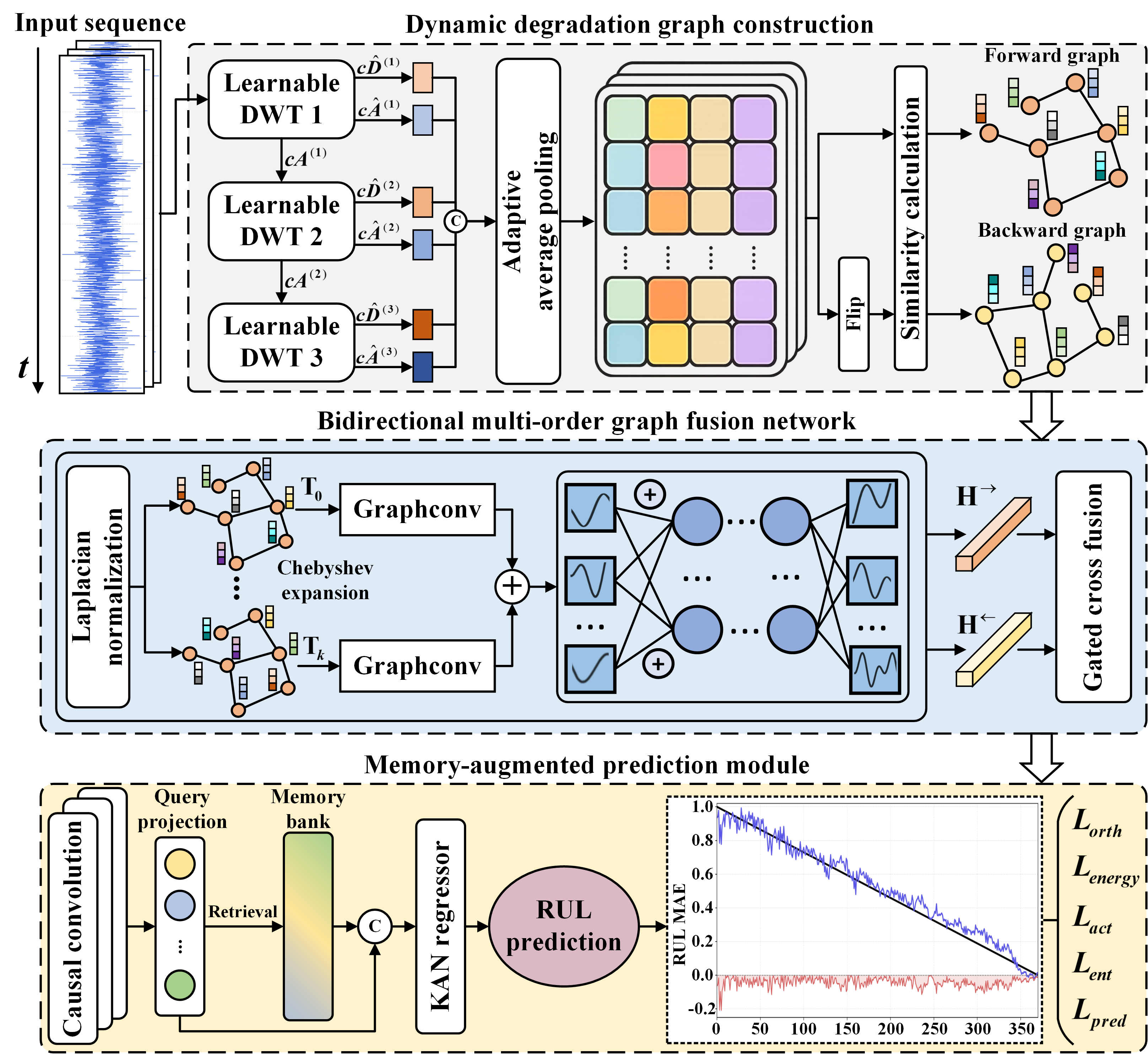}
	\caption{The overall architecture of our proposed method.}
	\label{FIG:1}
\end{figure}

In this section, we present the PE-BMGN framework for bearing RUL prediction. As shown in Figure \ref{FIG:1}, the proposed method aims to improve degradation-correlation modeling and prediction interpretability within a unified pipeline. Specifically, the overall framework consists of five stages: 1) the raw vibration sequence is first processed by a learnable discrete wavelet decomposition module to generate multi-scale degradation representations; 2) the decomposed features are then used to construct an adaptive graph, so that the dynamic correlations among different degradation states can be explicitly characterized; 3) based on the constructed graph, a bidirectional multi-order graph fusion network is developed to capture evolving local-to-global degradation dependencies by combining Chebyshev spectral graph convolution with Kolmogorov-Arnold nonlinear mapping; 4) a memory-augmented prediction module is further introduced to retrieve relevant historical degradation prototypes and perform final RUL regression through a Kolmogorov-Arnold regressor; and 5) a physically enhanced dynamic loss is developed by jointly considering the prediction objective, physical constraints, and Kolmogorov-Arnold regularization, thereby improving the robustness, physical consistency, and interpretability of the proposed framework.

\subsection{Dynamic degradation graph construction}

As shown in Figure~\ref{FIG:1}, dynamic degradation graph construction serves as the front-end representation module of the proposed PE-BMGN framework. Its objective is to transform raw vibration sequences into a graph structure in which each node represents a degradation state and each edge describes the correlation between two degradation states. This step is essential because bearing vibration signals are strongly non-stationary during the whole life cycle. Slow degradation tendencies, local impulsive responses, and noise-induced disturbances are usually mixed in the raw sequence. If graph topology is directly constructed from such raw observations, the resulting node similarities may be dominated by irrelevant amplitude fluctuations rather than degradation-related dependencies. To alleviate this problem, the proposed module first extracts adaptive multi-scale representations through learnable discrete wavelet decomposition, and then constructs a sparse adaptive graph based on the learned node representations.

\subsubsection{Learnable discrete wavelet decomposition}

\begin{figure}
	\centering
	\includegraphics[width=.8\textwidth]{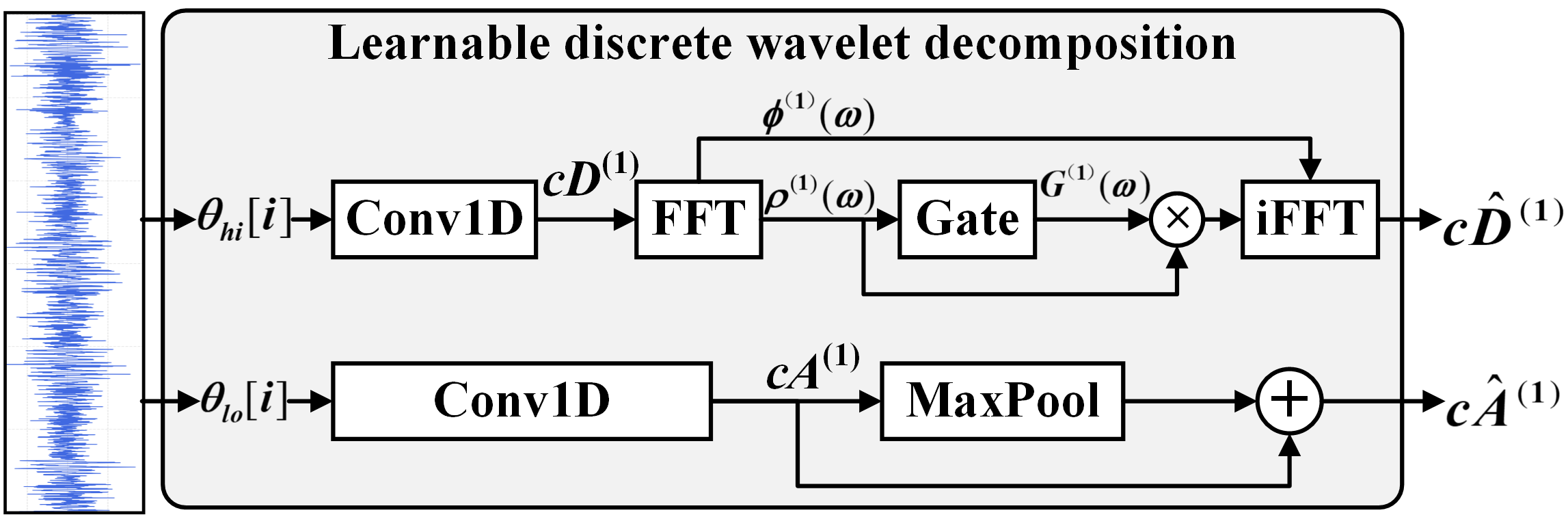}
	\caption{The first-level learnable discrete wavelet decomposition process. The same operations are recursively applied to subsequent levels, with the low-frequency output from the previous level serving as the input of the next level.}
	\label{FIG:2}
\end{figure}

Let $\mathbf{S}^{(0)}=\mathcal{X}$ be the input of the decomposition module, and let $\mathbf{S}^{(\ell-1)}$ be the input sequence at the $\ell$-th decomposition level. A learnable discrete wavelet decomposition module is introduced, as shown in Figure~\ref{FIG:2}. Different from conventional discrete wavelet transform with fixed bases \cite{heil1989continuous}, the decomposition filters in this module are trainable and optimized together with the downstream RUL prediction objective. Therefore, the decomposition process can adapt to different bearings and degradation stages instead of relying on a fixed signal basis.

At the $\ell$-th decomposition level, two branch coefficients are generated by one-dimensional convolution with learnable filters. The first branch is formulated as
\begin{equation}
{cD}^{(\ell)}
=
\operatorname{Conv1D}\!\left(\mathbf{S}^{(\ell-1)},\boldsymbol{\theta}_{lo}^{(\ell)}\right).
\end{equation}

The second branch is formulated as
\begin{equation}
{cA}^{(\ell)}
=
\operatorname{Conv1D}\!\left(\mathbf{S}^{(\ell-1)},\boldsymbol{\theta}_{hi}^{(\ell)}\right).
\end{equation}
where $\boldsymbol{\theta}_{lo}^{(\ell)}$ and $\boldsymbol{\theta}_{hi}^{(\ell)}$ denote the learnable decomposition filters at this level. Through these two trainable convolutional branches, the original sequence is projected into complementary coefficient spaces, which provides a basis for separating oscillatory degradation information and dominant temporal responses.

For the ${cD}^{(\ell)}$ branch, a frequency-domain gated enhancement strategy is adopted to suppress redundant oscillatory interference while preserving degradation-sensitive spectral components. Specifically, ${cD}^{(\ell)}$ is first transformed into the frequency domain, and its spectrum is decomposed into magnitude and phase components:
\begin{equation}
\mathcal{F}\!\left({cD}^{(\ell)}\right)(\omega)
=
\boldsymbol{\rho}^{(\ell)}(\omega)\exp\!\left(\mathrm{i}\boldsymbol{\phi}^{(\ell)}(\omega)\right).
\end{equation}
where $\mathcal{F}(\cdot)$ denotes the Fourier transform, $\omega$ is the frequency variable, $\mathrm{i}$ is the imaginary unit, and $\boldsymbol{\rho}^{(\ell)}(\omega)$ and $\boldsymbol{\phi}^{(\ell)}(\omega)$ are the spectral magnitude and phase, respectively. A learnable gate $\mathbf{G}^{(\ell)}(\omega)$ is then used to recalibrate the magnitude term, while the phase term is retained to avoid destroying the temporal structure of the original signal. The enhanced coefficient is obtained as
\begin{equation}
c\hat{D}^{(\ell)}
=
\mathcal{F}^{-1}\!\left(
\mathbf{G}^{(\ell)}(\omega)
\odot
\boldsymbol{\rho}^{(\ell)}(\omega)
\exp\!\left(\mathrm{i}\boldsymbol{\phi}^{(\ell)}(\omega)\right)
\right).
\end{equation}
where $\mathcal{F}^{-1}(\cdot)$ denotes the inverse Fourier transform and $\odot$ denotes element-wise multiplication. This operation enables the model to emphasize frequency components that are more closely related to degradation evolution while weakening irrelevant spectral fluctuations.

For the ${cA}^{(\ell)}$ branch, a max-pooling-based residual enhancement strategy is employed to strengthen dominant responses without discarding the original coefficient information. The enhanced coefficient is defined as
\begin{equation}
c\hat{A}^{(\ell)}
=
{cA}^{(\ell)}
+
\operatorname{MaxPool}\!\left({cA}^{(\ell)}\right).
\end{equation}

Compared with directly replacing the branch coefficient by a pooled representation, the residual form preserves the original temporal information and highlights salient local responses at the same time. In this way, the two branches provide complementary representations: the gated frequency branch focuses on spectral degradation signatures, whereas the residual pooling branch enhances dominant temporal responses.

\subsubsection{Dynamic graph topology construction}

\begin{figure}
	\centering
	\includegraphics[width=.8\textwidth]{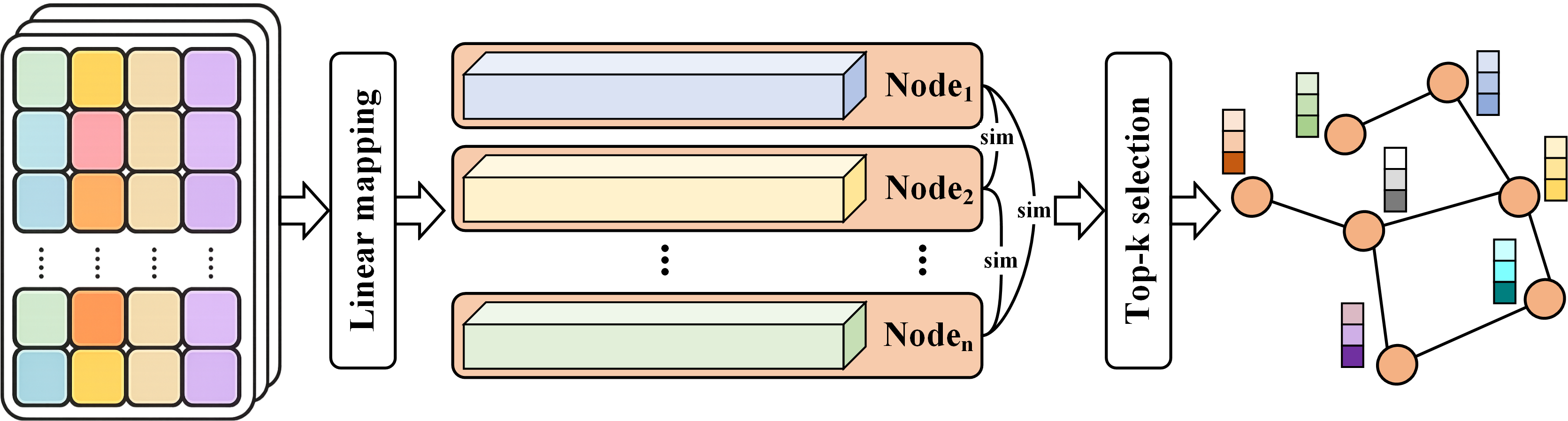}
	\caption{Dynamic graph construction process.}
	\label{FIG:3_1}
\end{figure}

After the multi-scale coefficients have been obtained, the next step is to generate node-wise degradation representations for graph construction. As shown in Figure~\ref{FIG:3_1}, the enhanced coefficients from different decomposition levels are compressed and aggregated into compact node features. The level-wise fused representation is first written as
\begin{equation}
\mathbf{u}_{t}^{(\ell)}
=
\hat{\mathbf{c}}_{A,t}^{(\ell)}
\oplus
\hat{\mathbf{c}}_{D,t}^{(\ell)}.
\end{equation}
where $\oplus$ denotes feature concatenation. Then, the multi-order representations are aggregated by adaptive average pooling:
\begin{equation}
\mathbf{x}_{t}
=
\mathcal{P}\!\left(
\mathbf{u}_{t}^{(1)}
\oplus
\mathbf{u}_{t}^{(2)}
\oplus
\cdots
\oplus
\mathbf{u}_{t}^{(J)}
\right).
\end{equation}
where $\mathcal{P}(\cdot)$ denotes adaptive average pooling. Accordingly, the node feature matrix of the degradation graph is obtained as
\begin{equation}
\mathbf{X}
=
\left[
\mathbf{x}_{1},
\mathbf{x}_{2},
\ldots,
\mathbf{x}_{T}
\right]^{\top}.
\end{equation}

Each row of $\mathbf{X}$ represents a degradation state containing both temporal and spectral information from multiple decomposition levels.

Based on the obtained node representations, an adaptive adjacency matrix is constructed to quantify pairwise degradation correlations. Two learnable projection matrices are first used to map node features into query and key spaces. The query representation of the $i$-th node is defined as
\begin{equation}
\mathbf{q}_{i}
=
\mathbf{W}_{q}\mathbf{x}_{i}.
\end{equation}

The key representation of the $j$-th node is defined as
\begin{equation}
\mathbf{k}_{j}
=
\mathbf{W}_{k}\mathbf{x}_{j}.
\end{equation}
where $\mathbf{W}_{q}$ and $\mathbf{W}_{k}$ are trainable projection matrices. The similarity between two degradation states is measured by cosine similarity:
\begin{equation}
s_{ij}
=
\frac{
\mathbf{q}_{i}^{\top}\mathbf{k}_{j}
}{
\left\|\mathbf{q}_{i}\right\|
\left\|\mathbf{k}_{j}\right\|
}.
\end{equation}

To control the sharpness of the similarity distribution, a temperature coefficient $\tau$ is introduced. The dense adaptive adjacency matrix is computed as
\begin{equation}
\tilde{A}_{ij}
=
\frac{
\exp\!\left(s_{ij}/\tau\right)
}{
\sum_{r=1}^{T}
\exp\!\left(s_{ir}/\tau\right)
}.
\end{equation}

A fully connected graph may introduce redundant weak dependencies and increase the difficulty of subsequent graph propagation. Therefore, a top-$K_{\mathrm{top}}$ sparsification strategy is further applied to retain only the most relevant neighbors for each node:
\begin{equation}
\mathcal{N}_{K_{\mathrm{top}}}(i)
=
\operatorname{TopK}_{K_{\mathrm{top}}}
\left(
\left\{
\tilde{A}_{ij}
\right\}_{j=1}^{T}
\right).
\end{equation}

The final sparse adaptive adjacency matrix is formulated as
\begin{equation}
A_{ij}
=
\frac{
\tilde{A}_{ij}\mathbf{1}_{\{j\in\mathcal{N}_{K_{\mathrm{top}}}(i)\}}
}{
\sum_{r\in\mathcal{N}_{K_{\mathrm{top}}}(i)}
\tilde{A}_{ir}
}.
\end{equation}
where $\mathbf{1}_{\{\cdot\}}$ is the indicator function. Through this sparsification and normalization process, weak and redundant connections are suppressed, while degradation states with strong correlations are preserved. Finally, the constructed graph can be represented as $\mathcal{G}=(\mathcal{V},\mathcal{E},\mathbf{X})$, with $\mathbf{A}$ describing its adaptive topology. The obtained $\mathbf{X}$ and $\mathbf{A}$ are then used as the inputs of the bidirectional multi-order graph fusion network, enabling the subsequent graph propagation process to capture local-to-global degradation dependencies along the bearing life cycle.

\subsection{Bidirectional multi-order graph fusion network}

After the adaptive degradation graph has been constructed, the next step is to extract degradation dependencies from the learned graph topology. As illustrated in Figure~\ref{FIG:1}, the proposed PE-BMGN framework employs a bidirectional Chebyshev graph Kolmogorov-Arnold network to propagate degradation information before memory-augmented prediction. This design is motivated by two characteristics of bearing degradation. First, the degradation process is progressive and stage-dependent, where early weak-fault patterns, middle-stage transition responses, and late-stage severe degradation signatures are strongly correlated along the life cycle. Second, the adaptive graph provides pairwise degradation correlations among temporal states, but a single propagation direction may not be sufficient to represent both historical accumulation and reverse structural calibration. Therefore, the proposed network extracts forward and backward graph representations and then adaptively integrates them through a gated cross fusion mechanism, as shown in Figure~\ref{FIG:3}.

\subsubsection{Bidirectional Chebyshev graph Kolmogorov-Arnold network}

Using $\mathbf{X}$ and $\mathbf{A}$ obtained from the dynamic degradation graph construction module, the learned topology is decomposed into a forward graph and a backward graph to model degradation dependencies from complementary directions. The forward graph and the backward graph are defined as
\begin{equation}
\mathbf{A}^{\rightarrow}
=
\mathbf{A},\qquad
   \mathbf{A}^{\leftarrow}
=
\mathbf{A}^{\top}.
\end{equation}
where $\mathbf{A}^{\rightarrow}$ preserves the original degradation correlation direction, and $\mathbf{A}^{\leftarrow}$ describes the reverse structural dependency. The forward branch emphasizes the accumulation of degradation information from earlier states to later states, whereas the backward branch provides reverse calibration cues from later degradation states to earlier weak-fault states. In this way, the complete degradation trajectory is represented from two complementary graph views rather than from a single unidirectional propagation path.

For each directional graph $\mathbf{A}^{(\xi)}$, where $\xi\in\{\rightarrow,\leftarrow\}$, Chebyshev spectral graph convolution is introduced to capture multi-order degradation dependencies. The diagonal degree matrix is first computed as
\begin{equation}
D^{(\xi)}_{ii}
=
\sum_{j=1}^{T}
A^{(\xi)}_{ij}.
\end{equation}

Then, the random-walk graph Laplacian is formulated as
\begin{equation}
\mathbf{L}_{g}^{(\xi)}
=
\mathbf{I}_{T}
-
\left(
\mathbf{D}^{(\xi)}
+
\epsilon\mathbf{I}_{T}
\right)^{-1}
\mathbf{A}^{(\xi)}.
\end{equation}
where $\mathbf{I}_{T}$ is the identity matrix and $\epsilon$ is a small constant for numerical stability. To ensure stable spectral approximation, the Laplacian is further scaled as
\begin{equation}
\tilde{\mathbf{L}}_{g}^{(\xi)}
=
\frac{2}{\lambda_{\max}^{(\xi)}}
\mathbf{L}_{g}^{(\xi)}
-
\mathbf{I}_{T}.
\end{equation}
where $\lambda_{\max}^{(\xi)}$ denotes the largest eigenvalue of $\mathbf{L}_{g}^{(\xi)}$. Based on the scaled Laplacian, Chebyshev basis responses are recursively generated. The corresponding propagation process is written as
\begin{equation}
\mathbf{T}_{0}^{(\xi)}
=
\mathbf{X}, \qquad
\mathbf{T}_{1}^{(\xi)}
=
\tilde{\mathbf{L}}_{g}^{(\xi)}
\mathbf{X}, \qquad
\mathbf{T}_{r}^{(\xi)}
=
2
\tilde{\mathbf{L}}_{g}^{(\xi)}
\mathbf{T}_{r-1}^{(\xi)}
-
\mathbf{T}_{r-2}^{(\xi)}.
\end{equation}
where $\mathbf{T}_{r}^{(\xi)}$ denotes the $r$-th order graph propagation response. In this way, the zeroth-order term preserves the original node state, the first-order term captures one-hop degradation interactions, and higher-order terms expand the receptive field to multi-hop correlated degradation states. This property is particularly suitable for bearing RUL prediction, because the RUL of one temporal state is usually influenced not only by its adjacent monitoring states, but also by distant states with similar degradation signatures.

Although Chebyshev graph convolution can capture local-to-global structural dependencies, conventional graph convolution usually applies linear transformations after graph aggregation. Such a design is insufficient to describe the nonlinear and stage-varying degradation behavior of bearings. Therefore, a Kolmogorov-Arnold nonlinear mapping is introduced after each Chebyshev propagation order. For the $r$-th order response in direction $\xi$, the nonlinear transformation is defined as
\begin{equation}
\Phi_{r}^{(\xi)}
\left(
\mathbf{T}_{r}^{(\xi)}
\right)
=
\mathbf{W}_{b,r}^{(\xi)}
\sigma
\left(
\mathbf{T}_{r}^{(\xi)}
\right)
+
\mathbf{W}_{sp,r}^{(\xi)}
\mathcal{B}
\left(
\mathbf{T}_{r}^{(\xi)}
\right).
\end{equation}
where $\sigma(\cdot)$ denotes the base activation function, $\mathcal{B}(\cdot)$ denotes the spline basis expansion, and $\mathbf{W}_{b,r}^{(\xi)}$ and $\mathbf{W}_{sp,r}^{(\xi)}$ are learnable coefficients. Compared with MLP-like graph transformations, this formulation represents nonlinear feature interactions through structured spline-based functions, which improves the ability of the model to describe complex degradation mappings while maintaining functional interpretability.

The output of one directional Chebyshev graph Kolmogorov-Arnold layer is obtained by aggregating all propagation orders:
\begin{equation}
\mathbf{Z}^{(\xi)}
=
\sum_{r=0}^{K_c-1}
\Phi_{r}^{(\xi)}
\left(
\mathbf{T}_{r}^{(\xi)}
\right)
+
\mathbf{b}^{(\xi)}.
\end{equation}
where $K_c$ is the Chebyshev order and $\mathbf{b}^{(\xi)}$ is the bias term. To improve representation stability, normalization and activation operations are further applied:
\begin{equation}
\mathbf{H}^{(\xi)}
=
\delta
\left(
\operatorname{Norm}
\left(
\mathbf{Z}^{(\xi)}
\right)
\right).
\end{equation}
where $\operatorname{Norm}(\cdot)$ denotes the normalization operation, and $\delta(\cdot)$ denotes the nonlinear activation function used after normalization. Accordingly, the forward and backward degradation representations are obtained as
\begin{equation}
\mathbf{H}^{\rightarrow}
=
f_{\mathrm{CGKAN}}
\left(
\mathbf{X},
\mathbf{A}^{\rightarrow}
\right), \qquad
\mathbf{H}^{\leftarrow}
=
f_{\mathrm{CGKAN}}
\left(
\mathbf{X},
\mathbf{A}^{\leftarrow}
\right).
\end{equation}
where $f_{\mathrm{CGKAN}}(\cdot)$ denotes the Chebyshev graph Kolmogorov-Arnold propagation operator. The two directional representations encode complementary degradation information: $\mathbf{H}^{\rightarrow}$ focuses on progressive degradation accumulation, while $\mathbf{H}^{\leftarrow}$ reflects reverse structural calibration along the life trajectory.

\subsubsection{Gated cross fusion module}

\begin{figure}
	\centering
	\includegraphics[width=.8\textwidth]{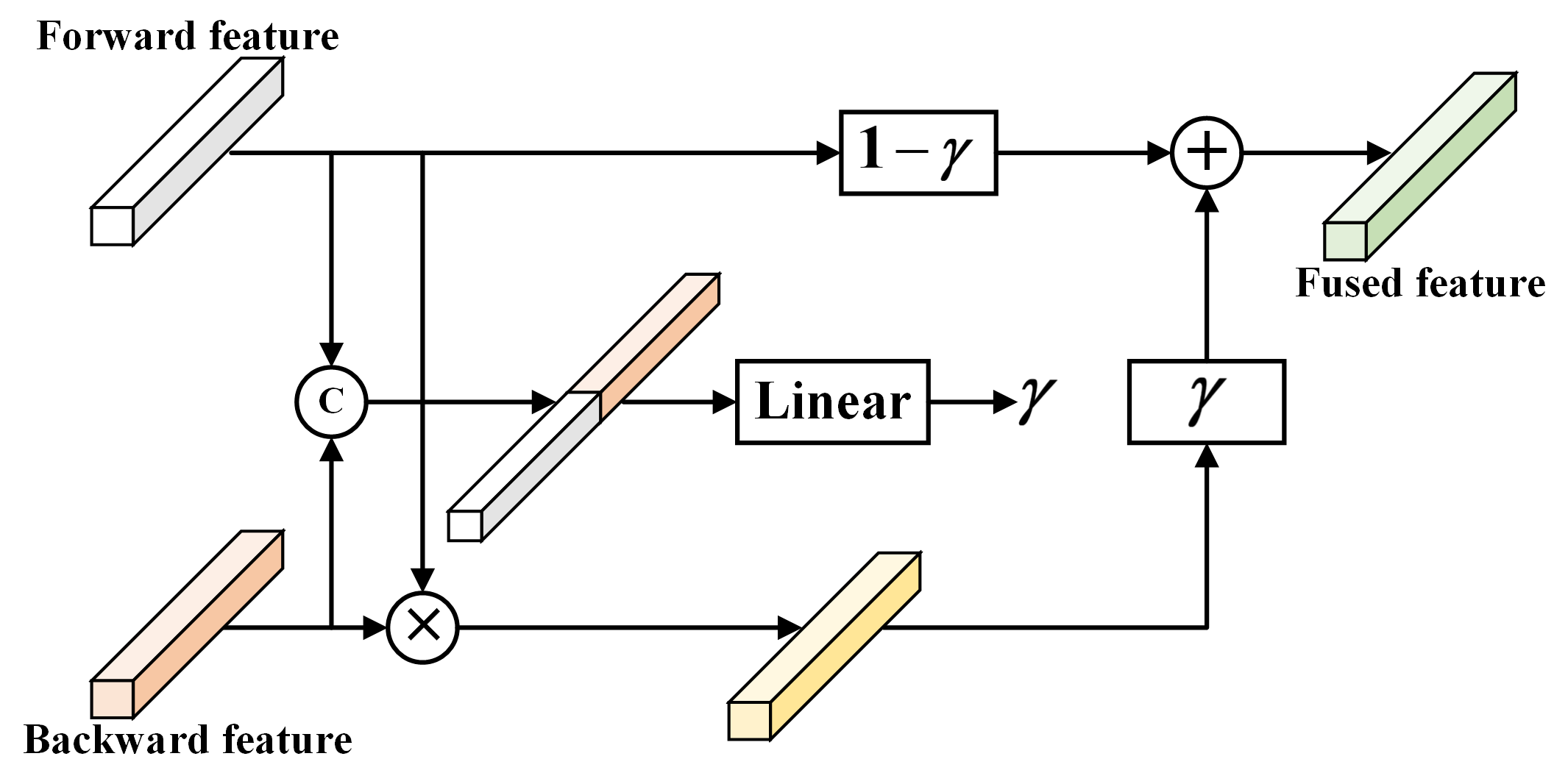}
	\caption{Gated cross fusion module.}
	\label{FIG:3}
\end{figure}

After obtaining the forward and backward graph representations, a gated cross fusion mechanism is introduced to adaptively integrate the two directional features, as shown in Figure~\ref{FIG:3}. Direct concatenation treats the two directions as equally important, which is inconsistent with the stage-varying nature of bearing degradation. In early degradation stages, the backward branch can provide useful structural cues for calibrating weak degradation signatures. In contrast, when the bearing approaches severe degradation, the forward branch may become more important because it contains accumulated degradation information from previous states. Therefore, a learnable gate is used to dynamically control the contribution of the two directional branches.

To generate the gate, the forward feature, the backward feature, and their element-wise interaction are first combined:
\begin{equation}
\mathbf{R}
=
\left[
\mathbf{H}^{\rightarrow}
\|
\mathbf{H}^{\leftarrow}
\|
\left(
\mathbf{H}^{\rightarrow}
\odot
\mathbf{H}^{\leftarrow}
\right)
\right].
\end{equation}
where $\|$ denotes feature concatenation. The interaction term explicitly describes the consistency between the two directional features, which helps the fusion module distinguish complementary information from redundant responses. The gate coefficient is then calculated as
\begin{equation}
\boldsymbol{\gamma}
=
\operatorname{Sigmoid}
\left(
\mathbf{W}_{\gamma}
\mathbf{R}
+
\mathbf{b}_{\gamma}
\right).
\end{equation}
where $\boldsymbol{\gamma}$ is the adaptive fusion gate, and $\mathbf{W}_{\gamma}$ and $\mathbf{b}_{\gamma}$ are learnable parameters. The final fused degradation representation is obtained by
\begin{equation}
\mathbf{H}^{\mathrm{fuse}}
=
\left(
1
-
\boldsymbol{\gamma}
\right)
\odot
\mathbf{H}^{\rightarrow}
+
\boldsymbol{\gamma}
\odot
\mathbf{H}^{\leftarrow}.
\end{equation}

Finally, the fused representation is denoted as
\begin{equation}
\mathbf{H}^{\mathrm{fuse}}
=
\left[
\mathbf{h}_{1},
\mathbf{h}_{2},
\ldots,
\mathbf{h}_{T}
\right]^{\top}.
\end{equation}
where $\mathbf{h}_{t}\in\mathbb{R}^{d_h}$ is the degradation representation of the $t$-th temporal state after bidirectional graph propagation and gated cross fusion. The obtained representation is subsequently fed into the memory-augmented prediction network for degradation prototype retrieval and final RUL regression.

\subsection{Memory-augmented prediction module}

After the bidirectional multi-order graph fusion network, the fused representation contains dynamic degradation dependencies extracted from both forward and backward graph propagation. However, if the prediction module relies only on the current latent representation, useful historical degradation knowledge may be insufficiently exploited, which can weaken the model's ability to recognize similar degradation states and capture strongly nonlinear degradation trends. To address this issue, an interpretable memory-augmented prediction network is introduced to retrieve historical degradation prototypes and incorporate them into the final RUL regression stage.

Using the fused state $\mathbf{h}_{t}$, a dynamic memory bank is constructed as $\mathbf{M}=[\mathbf{m}_{1},\mathbf{m}_{2},\ldots,\mathbf{m}_{F}]^{\top}\in\mathbb{R}^{F\times d_m}$, where $F$ is the memory capacity and $d_m$ is the memory dimension. Each memory vector can be regarded as a historical degradation prototype that stores representative degradation information. To retrieve prototypes relevant to the current degradation state, $\mathbf{h}_{t}$ is first projected into a memory-query space:
\begin{equation}
\boldsymbol{\eta}_{t}
=
\mathbf{W}_{m}\mathbf{h}_{t}
+
\mathbf{b}_{m}.
\end{equation}
where $\boldsymbol{\eta}_{t}$ is the memory query of the $t$-th temporal state, and $\mathbf{W}_{m}$ and $\mathbf{b}_{m}$ are learnable projection parameters. The relevance between the current degradation state and the $j$-th memory prototype is measured by a softmax-normalized similarity score:
\begin{equation}
\pi_{t,j}
=
\frac{
\exp\left(\boldsymbol{\eta}_{t}^{\top}\mathbf{m}_{j}\right)
}{
\sum_{r=1}^{F}
\exp\left(\boldsymbol{\eta}_{t}^{\top}\mathbf{m}_{r}\right)
}.
\end{equation}
where $\pi_{t,j}$ denotes the relevance weight assigned to the $j$-th memory prototype. To avoid introducing unrelated historical patterns, only the top-$K_m$ prototypes with the highest relevance weights are selected:
\begin{equation}
\Gamma_{t}^{K_m}
=
\operatorname{TopK}_{K_m}
\left(
\left\{
\pi_{t,j}
\right\}_{j=1}^{F}
\right).
\end{equation}

The selected relevance weights are then renormalized within the retrieved prototype set:
\begin{equation}
\bar{\pi}_{t,j}
=
\frac{
\pi_{t,j}
}{
\sum_{r\in\Gamma_{t}^{K_m}}
\pi_{t,r}
}.
\end{equation}
where $j\in\Gamma_{t}^{K_m}$, and $\bar{\pi}_{t,j}$ represents the normalized contribution of the selected prototype. The memory-enhanced degradation representation is obtained by weighted prototype aggregation:
\begin{equation}
\mathbf{h}^{\mathrm{mem}}_{t}
=
\sum_{j\in\Gamma_{t}^{K_m}}
\bar{\pi}_{t,j}
\mathbf{m}_{j}.
\end{equation}

Through this weighted retrieval strategy, the memory module does not simply average the selected historical prototypes. Instead, it assigns larger contributions to prototypes that are more consistent with the current degradation state, enabling the prediction network to exploit historical degradation knowledge in a more discriminative and interpretable manner.

To maintain the adaptability of the memory bank during training, a momentum update strategy is adopted. Specifically, the memory prototypes are progressively updated by combining the previous memory states with the average memory query representation of the current mini-batch. This update strategy prevents abrupt changes in the memory prototypes and allows the memory bank to gradually absorb newly learned degradation information. As a result, the stored prototypes can remain stable while still adapting to the evolving feature distribution during model optimization.

After memory retrieval, the current memory query representation and the retrieved historical prototype representation are concatenated to form the prediction feature:
\begin{equation}
\mathbf{z}_{t}
=
\boldsymbol{\eta}_{t}
\,
\|
\,
\mathbf{h}^{mem}_{t}.
\end{equation}

The concatenated feature contains both the current degradation state and the retrieved historical degradation knowledge, thereby providing a more informative basis for final RUL regression.

Finally, a Kolmogorov-Arnold regressor is employed to map the memory-enhanced representation into the predicted RUL value. Specifically, the regressor is composed of Kolmogorov-Arnold nonlinear mapping layers, which replace conventional fully connected transformations with structured functional mappings. The prediction process can be written as
\begin{equation}
\hat{y}_{t}
=
\Psi_{2}
\left(
\delta
\left(
\Psi_{1}(\mathbf{z}_{t})
\right)
\right).
\end{equation}
where $\Psi_{1}(\cdot)$ and $\Psi_{2}(\cdot)$ denote two Kolmogorov-Arnold mapping layers. For a general Kolmogorov-Arnold mapping layer, the transformation can be represented as
\begin{equation}
\Psi(\mathbf{u})
=
\mathbf{W}_{b}^{\Psi}
\sigma(\mathbf{u})
+
\mathbf{W}_{sp}^{\Psi}
\mathcal{B}(\mathbf{u}).
\end{equation}
where $\mathbf{u}$ is the input of the mapping layer, and $\mathbf{W}_{b}^{\Psi}$ and $\mathbf{W}_{sp}^{\Psi}$ are learnable coefficients. This formulation enables the regressor to learn complex nonlinear relationships through spline-based functional mappings, which is more suitable for capturing the strong nonlinear degradation characteristics of bearings than a purely linear prediction head.

\subsection{Physics-enhanced dynamic loss function}

The proposed PE-BMGN framework contains several learnable components. If the model is optimized only by the prediction error, the learned representations may fit the training degradation trajectories but fail to preserve the physical consistency of signal decomposition or the stability of the nonlinear mappings. Therefore, a physics-enhanced dynamic loss function is developed to guide the optimization of the whole framework.

First, the prediction objective is defined to minimize the discrepancy between the predicted and ground-truth normalized RUL sequences:
\begin{equation}
\mathcal{L}_{pred}
=
\frac{1}{N}
\sum_{b=1}^{N}
\frac{1}{T_b}
\sum_{t=1}^{T_b}
\left(\hat{y}^{(b)}_{t}-y^{(b)}_{t}\right)^{2}.
\end{equation}

Second, physical consistency constraints are introduced for the learnable wavelet decomposition module. Since the decomposition filters are updated during training, their physical properties may gradually deviate from the desirable wavelet structure if no constraint is imposed. To alleviate this problem, orthogonality and energy constraints are considered. For the learnable filters introduced in Section~3.1, the orthogonality loss is defined as
\begin{equation}
\mathcal{L}_{orth}
=
\mu_{orth}
\sum_{\ell=1}^{J}
\left\|
\operatorname{Conv1D}
\left(
\boldsymbol{\theta}_{lo}^{(\ell)},
\operatorname{flip}\left(\boldsymbol{\theta}_{hi}^{(\ell)}\right)
\right)
-
\mathbf{e}_{0}
\right\|_{2}^{2}.
\end{equation}
where $\mu_{orth}$ is the weighting coefficient, $\operatorname{flip}(\cdot)$ denotes the reversed filter operation, and $\mathbf{e}_{0}$ denotes the ideal impulse response. This term encourages the learned low-pass and high-pass filters to maintain an approximately orthogonal relationship, thereby reducing redundant responses between the two decomposition branches.

Meanwhile, the energy loss is introduced to prevent the learnable filters from producing excessive amplification or attenuation during signal decomposition:
\begin{equation}
\mathcal{L}_{energy}
=
\mu_{energy}
\sum_{\ell=1}^{J}
\left[
\left(
\left\|
\boldsymbol{\theta}_{lo}^{(\ell)}
\right\|_{2}^{2}
-1
\right)^{2}
+
\left(
\left\|
\boldsymbol{\theta}_{hi}^{(\ell)}
\right\|_{2}^{2}
-1
\right)^{2}
\right].
\end{equation}
where $\mu_{energy}$ is the weighting coefficient. This term constrains the energy of the two learnable filters to remain close to a normalized scale. In this way, the front-end decomposition module can adapt to the downstream RUL prediction objective while still preserving the basic physical plausibility of wavelet-like signal decomposition.

Third, Kolmogorov-Arnold regularization terms are introduced to constrain the nonlinear mappings in both the bidirectional graph module and the prediction module. In the proposed framework, Kolmogorov-Arnold mappings are used to replace conventional linear or multilayer-perceptron transformations, which improves the ability to model complex degradation patterns. However, overly flexible spline-based mappings may also introduce redundant functional responses. Therefore, activation and entropy regularization terms are imposed on the spline coefficients to improve the stability and interpretability of nonlinear representation learning.

For the spline coefficients in KAN layers, the activation regularization term is formulated as
\begin{equation}
\mathcal{L}_{act}
=
\beta_{1}
\sum_{o=1}^{n_{out}}
\sum_{i=1}^{n_{in}}
\operatorname{mean}_{k}
\left|
W^{spline}_{o,i,k}
\right|.
\end{equation}
where $W^{spline}_{o,i,k}$ denotes the $k$-th spline coefficient between the $i$-th input channel and the $o$-th output channel, $n_{in}$ and $n_{out}$ denote the numbers of input and output channels, respectively, and $\beta_{1}$ is the regularization coefficient. This term penalizes excessively large spline activations and encourages the KAN layers to learn compact functional mappings, thereby suppressing redundant nonlinear responses.

To further regularize the distribution of active spline connections, the entropy regularization term is defined as
\begin{equation}
\mathcal{L}_{ent}
=
-\beta_{2}
\sum_{o,i}
p_{o,i}
\log
p_{o,i}.
\end{equation}
where $\beta_{2}$ is the corresponding regularization coefficient, and $p_{o,i}$ denotes the normalized importance of the spline connection between the $i$-th input channel and the $o$-th output channel. This term constrains the distribution of spline responses across different input-output pairs, preventing the KAN layers from relying on overly dispersed and unstable functional activations. Together with $\mathcal{L}_{act}$, it encourages the Kolmogorov-Arnold mappings to learn sparse, compact, and stable nonlinear transformations, thereby improving the interpretability of degradation representation learning.

Finally, the overall training objective is obtained by combining the prediction loss, the physical constraint losses, and the Kolmogorov-Arnold regularization losses:
\begin{equation}
\mathcal{L}_{total}
=
\mathcal{L}_{pred}
+
\mathcal{L}_{orth}
+
\mathcal{L}_{energy}
+
\mathcal{L}_{act}
+
\mathcal{L}_{ent}.
\end{equation}

Through this joint optimization objective, the proposed framework can simultaneously reduce RUL prediction errors, preserve physically meaningful decomposition properties, and constrain the complexity of KAN-based nonlinear mappings. As a result, the model is encouraged to learn degradation representations that are not only accurate for prediction, but also compact, stable, and interpretable from the perspective of functional transformation.

\section{Experimentation and analysis}

In this section, comprehensive experiments are conducted to evaluate the effectiveness of the proposed PE-BMGN. Section 4.1 provides a detailed description of the XJTU-SY and PHM2012 datasets. Section~4.2 introduces the evaluation metrics used for quantitative comparison. Section~4.3 describes the implementation details. Section~4.4 presents the experimental results and analysis on the XJTU-SY bearing dataset. Section~4.5 further evaluates the proposed method on the PHM2012 bearing dataset. Section~4.6 conducts ablation experiments to investigate the contributions of key components and hyperparameters in the proposed framework.

\subsection{Dataset description}

The XJTU-SY bearing dataset \cite{wang2018hybrid} is a benchmark for evaluating the proposed RUL prediction method. This dataset provides complete degradation trajectories of rolling bearings collected from accelerated life tests, as illustrated in Figure \ref{fig:xjtu_sy_platform}. It contains 15 bearings under three operating conditions with different speed-load combinations, including 2100 rpm/12 kN, 2250 rpm/11 kN, and 2400 rpm/10 kN. For each bearing, vibration signals in the horizontal and vertical directions are recorded by two accelerometers, and each monitoring sample consists of 32768 points. Since all bearing runs start from a healthy state and end at failure, the dataset is suitable for evaluating whether a model can capture nonlinear degradation evolution under different operating conditions.

\begin{center}
    \includegraphics[width=0.85\textwidth]{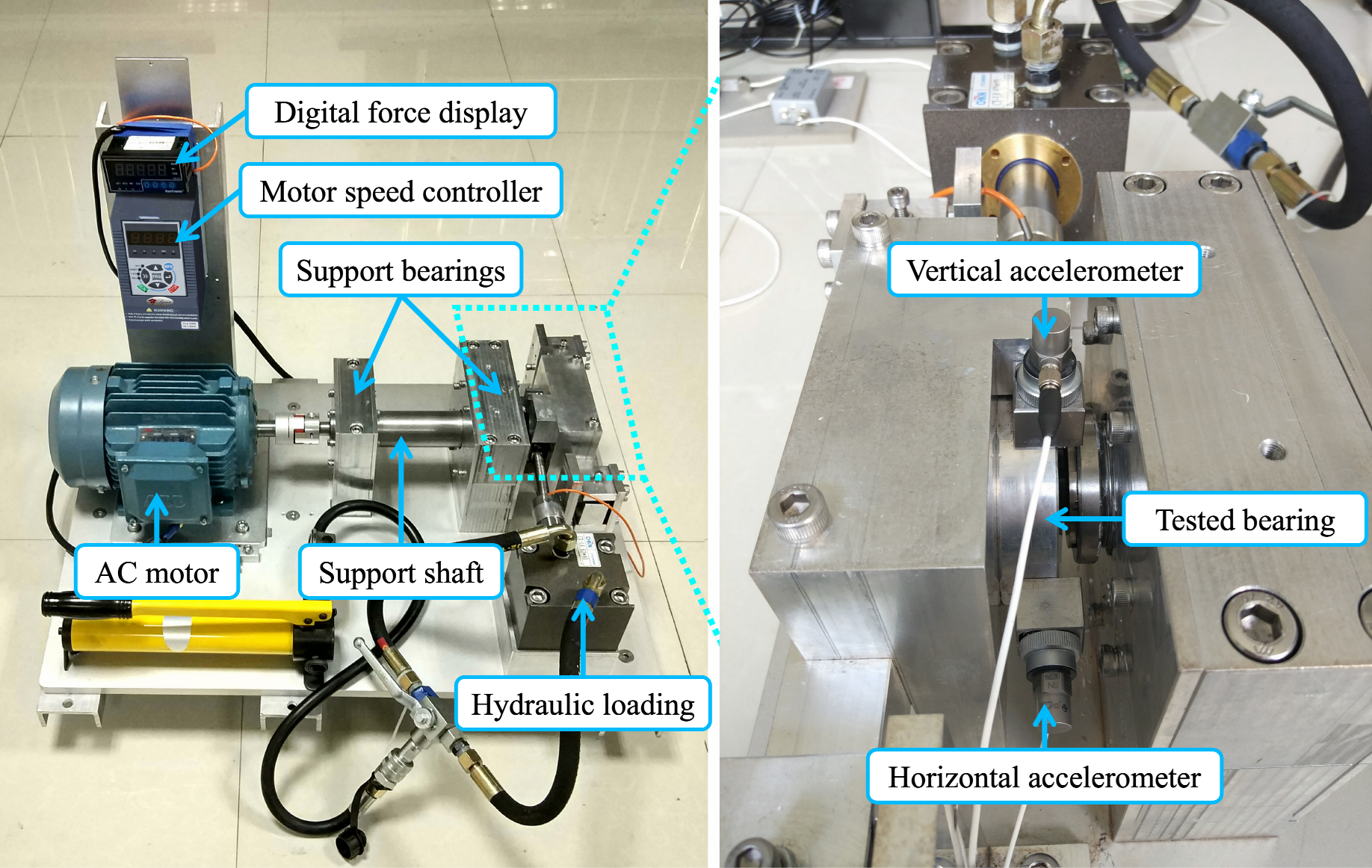}
    \captionof{figure}{XJTU-SY bearing test bench.}
    \label{fig:xjtu_sy_platform}
\end{center}

The PHM2012 bearing dataset \cite{nectoux2012pronostia} is adopted as the second benchmark for evaluating the proposed RUL prediction method. This dataset provides complete degradation trajectories of rolling bearings collected under accelerated life tests on the PRONOSTIA platform, as illustrated in Figure \ref{fig:phm2012_platform}. It contains three operating conditions with different speed-load combinations, including 1800 rpm/4000 N, 1650 rpm/4200 N, and 1500 rpm/5000 N. For each bearing, vibration signals in the horizontal and vertical directions are recorded, and each monitoring sample consists of 2560 points. Since all bearing runs start from a healthy state and end at failure, the dataset is suitable for evaluating whether a model can capture long-term degradation evolution under different operating conditions.

\begin{center}
    \includegraphics[width=0.85\textwidth]{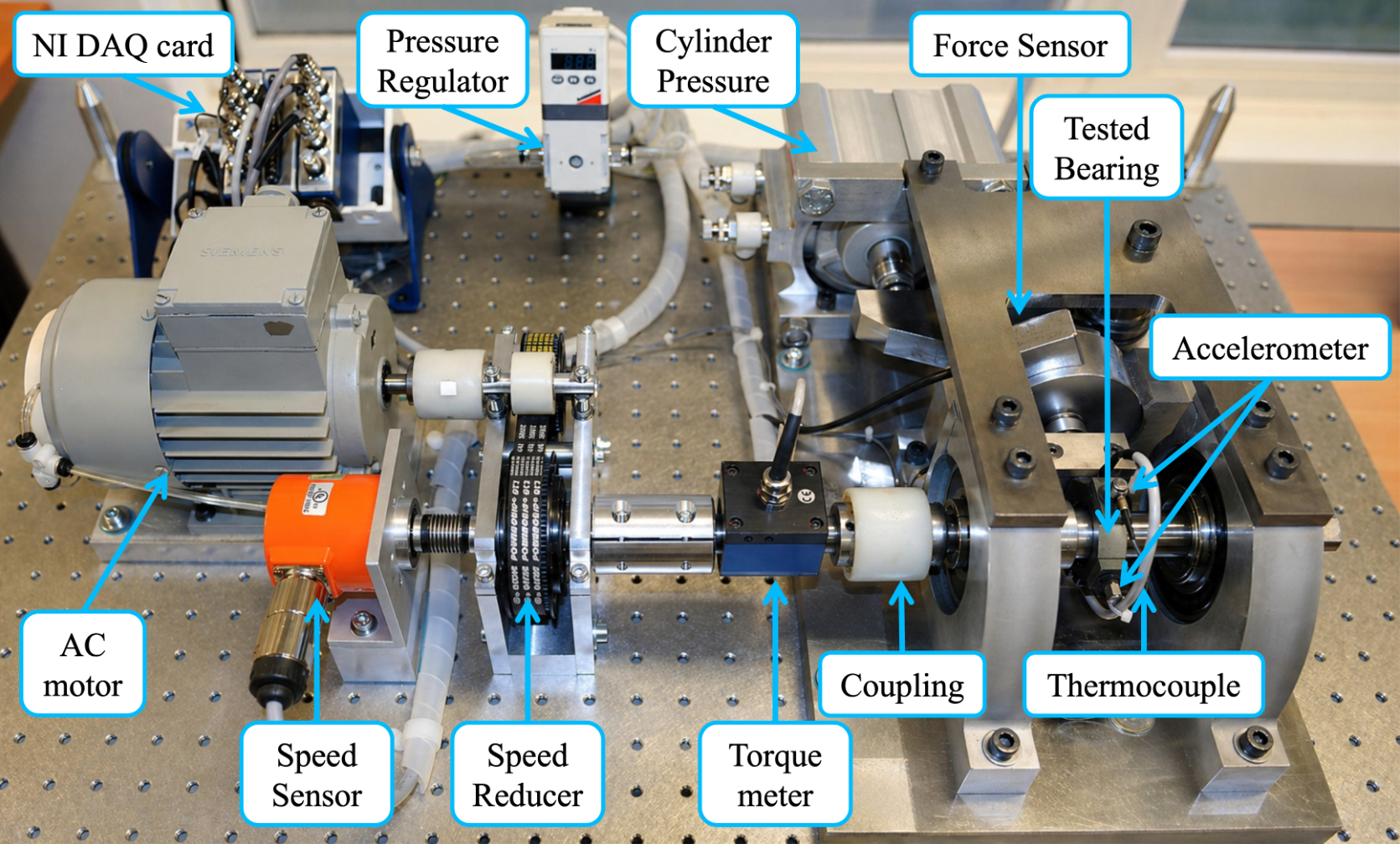}
    \captionof{figure}{PRONOSTIA bearing test bench.}
    \label{fig:phm2012_platform}
\end{center}

\subsection{Evaluation metrics}

To evaluate the RUL prediction performance, three commonly used metrics are adopted, namely root mean square error (RMSE), mean absolute error (MAE), and engineering application score (EAS). RMSE and MAE measure the numerical deviation between the predicted RUL and the ground truth, while EAS further considers the practical risk caused by early or late prediction in engineering maintenance. Let $y_t$ and $\hat{y}_t$ denote the true and predicted RUL values of the $t$-th sample, respectively, and let $T$ be the number of test samples. The RMSE and MAE are defined as
\begin{equation}
\mathrm{RMSE}
=
\sqrt{
\frac{1}{T}
\sum_{t=1}^{T}
\left(\hat{y}_{t}-y_{t}\right)^{2}
},
\end{equation}

\begin{equation}
\mathrm{MAE}
=
\frac{1}{T}
\sum_{t=1}^{T}
\left|\hat{y}_{t}-y_{t}\right|.
\end{equation}

The EAS is calculated as
\begin{equation}
\mathrm{EAS}
=
\frac{1}{T}
\sum_{t=1}^{T}
s_t,
\qquad
s_t=
\begin{cases}
\exp\left(\dfrac{y_t-\hat{y}_t}{13}\right)-1, & y_t>\hat{y}_t, \\
\exp\left(\dfrac{\hat{y}_t-y_t}{10}\right)-1, & y_t\leq \hat{y}_t .
\end{cases}
\end{equation}

For all three metrics, smaller values indicate better prediction performance.

\subsection{Implementation details}

The proposed PE-BMGN is implemented based on PyTorch \cite{paszke2019pytorch}. For the XJTU-SY dataset, the leave-one-out setting is adopted within each operating condition. For the PHM2012 dataset, the original challenge protocol is followed. Under Condition I, Bearing1\_1 and Bearing1\_2 are used for training, while Bearing1\_3 to Bearing1\_7 are used for testing. Under Condition II, Bearing2\_1 and Bearing2\_2 are used for training, while Bearing2\_3 to Bearing2\_7 are used for testing. Under Condition III, Bearing3\_1 and Bearing3\_2 are used for training, while Bearing3\_3 is used for testing. 

For the model configuration, the learnable discrete wavelet decomposition module contains three decomposition layers, and the wavelet filters are initialized by DB4 coefficients with a kernel size of 8. The dimension of the pooled multi-scale node representation is set to 200 for the XJTU-SY dataset and 80 for the PHM2012 dataset. In the dynamic graph construction module, the graph projection dimension is set to 6. The temperature coefficient is set to 0.6 for XJTU-SY and 0.5 for PHM2012. To construct a sparse adaptive graph, the top-160 most relevant neighbors are retained for each node on the XJTU-SY dataset, while the top-110 most relevant neighbors are retained on the PHM2012 dataset. In the bidirectional multi-order graph fusion network, two layers are used, and their Chebyshev orders are set to 2 and 3, respectively. Random-walk normalization is adopted for graph Laplacian normalization. The dynamic memory bank contains 128 memory prototypes for XJTU-SY and 64 memory prototypes for PHM2012, with the memory dimension fixed to 64. During memory-augmented prediction, the top-5 most relevant memory prototypes are retrieved. The Adam optimizer \cite{kingma2014adam} is adopted with an initial learning rate of 0.001. The maximum number of training epochs is set to 300. The temporal convolution module contains five layers with a kernel size of 5.

\subsection{Case study 1: XJTU-SY bearing dataset}

\subsubsection{Comparative experiments}

To validate the superiority of the our proposed method, 11 state-of-the-art RUL prediction methods are selected for comparative analysis in this paper, including T-GCN \cite{yang2022bearing}, ChebGCN-LSTM \cite{yang2022node}, SAGCN-SA \cite{wei2023bearing}, MR-LSTM \cite{xu2023multi}, SDFEPN \cite{yang2023dual}, CBAM-MSCNN \cite{du2022remaining}, DyWave-BiAGCN \cite{zhao2025new}, DC-DGCN \cite{he2025prediction}, GCN \cite{kipf2016semi}, STACP-GCN \cite{zhao2025node}, CNN-LSTM \cite{kim2019predicting} and SSPGKAN \cite{shen2025sparse}.

Table~\ref{tab:rmse_xjtu} reports the RMSE comparison results on the XJTU-SY dataset. PE-BMGN achieves the lowest Total\_Avg RMSE of 0.069, clearly outperforming the strongest baseline DyWave-BiAGCN with 0.096 and the KAN-based baseline SSPGKAN with 0.224. Since RMSE is more sensitive to large prediction deviations, this result indicates that the proposed method can more effectively suppress severe RUL estimation errors across different bearing trajectories. Compared with DyWave-BiAGCN, PE-BMGN further improves the overall RMSE by introducing bidirectional multi-order graph modeling into the graph propagation process. This advantage is particularly evident under Condition I and Condition II, where PE-BMGN obtains the best C1\_Avg and C2\_Avg RMSE values of 0.025 and 0.068, respectively, while DyWave-BiAGCN obtains 0.096 and 0.086. Although the C3\_Avg RMSE of PE-BMGN is 0.114, slightly higher than DyWave-BiAGCN with 0.106 due to the large error on Br3\_1, it is still much lower than SSPGKAN with 0.279. This comparison suggests that simply introducing KAN does not necessarily guarantee strong RUL prediction performance. SSPGKAN relies on explicit degradation stage division and stage-specific path graph construction, which may fragment the continuous degradation trajectory and make the prediction sensitive to stage boundary estimation. In contrast, PE-BMGN embeds KAN into bidirectional multi-order graph propagation and performs gated fusion of forward and backward degradation dependencies, enabling more continuous and discriminative modeling of local-to-global degradation correlations.

\begin{center}
\captionof{table}{Comparison results of RMSE on the XJTU-SY dataset. ``Br'' denotes a specific bearing sample, and ``C1'', ``C2'', and ``C3'' denote different operating conditions, where C1\_Avg, C2\_Avg, and C3\_Avg represent the average results of all test bearings under the corresponding condition.}
\label{tab:rmse_xjtu}
\scriptsize
\resizebox{\textwidth}{!}{
\begin{tabular}{lccccccccccccccccccc}
\hline
Method & Br1\_1 & Br1\_2 & Br1\_3 & Br1\_4 & Br1\_5 & C1\_Avg & Br2\_1 & Br2\_2 & Br2\_3 & Br2\_4 & Br2\_5 & C2\_Avg & Br3\_1 & Br3\_2 & Br3\_3 & Br3\_4 & Br3\_5 & C3\_Avg & Total\_Avg \\
\hline
T-GCN & 0.116 & 0.099 & 0.152 & 0.141 & 0.163 & 0.134 & 0.126 & 0.124 & 0.111 & 0.125 & 0.136 & 0.124 & 0.144 & 0.185 & 0.154 & 0.124 & 0.208 & 0.163 & 0.141 \\
ChebGCN-LSTM & 0.112 & 0.147 & 0.154 & 0.163 & 0.178 & 0.151 & 0.134 & 0.110 & 0.152 & 0.155 & 0.127 & 0.136 & 0.165 & 0.151 & 0.106 & 0.122 & 0.183 & 0.145 & 0.144 \\
SAGCN-SA & 0.106 & 0.115 & 0.124 & 0.090 & 0.173 & 0.122 & 0.125 & 0.116 & 0.121 & 0.127 & 0.102 & 0.118 & 0.159 & 0.164 & 0.115 & 0.120 & 0.187 & 0.149 & 0.130 \\
MR-LSTM & 0.097 & 0.108 & 0.140 & 0.071 & 0.120 & 0.107 & 0.092 & 0.100 & 0.096 & 0.095 & 0.117 & 0.100 & 0.152 & 0.163 & 0.101 & 0.099 & 0.195 & 0.142 & 0.116 \\
SDFEPN & 0.089 & 0.099 & 0.114 & 0.100 & 0.154 & 0.111 & 0.117 & 0.080 & 0.128 & 0.086 & 0.102 & 0.103 & \textbf{0.103} & 0.155 & 0.066 & 0.090 & 0.173 & 0.117 & 0.110 \\
CBAM-MSCNN & 0.092 & 0.102 & 0.138 & 0.106 & 0.160 & 0.120 & \textbf{0.085} & 0.097 & 0.132 & 0.091 & 0.112 & 0.103 & 0.144 & \textbf{0.110} & 0.098 & \textbf{0.057} & 0.165 & 0.115 & 0.113 \\
DyWave-BiAGCN & 0.087 & 0.072 & 0.119 & 0.078 & 0.125 & 0.096 & 0.096 & 0.085 & 0.102 & 0.069 & 0.079 & 0.086 & 0.113 & 0.117 & 0.070 & 0.061 & 0.169 & \textbf{0.106} & 0.096 \\
DC-DGCN & 0.105 & 0.067 & 0.076 & 0.125 & 0.148 & 0.104 & 0.120 & 0.092 & 0.129 & 0.096 & 0.091 & 0.106 & 0.106 & 0.136 & 0.129 & 0.109 & 0.184 & 0.133 & 0.114 \\
GCN & 0.140 & 0.135 & 0.139 & 0.119 & 0.180 & 0.143 & 0.152 & 0.160 & 0.140 & 0.151 & 0.136 & 0.148 & 0.188 & 0.232 & 0.195 & 0.153 & 0.249 & 0.203 & 0.165 \\
STACP-GCN & 0.074 & 0.068 & 0.089 & 0.100 & 0.135 & 0.093 & 0.104 & 0.087 & 0.105 & 0.104 & 0.077 & 0.095 & 0.113 & 0.098 & 0.084 & 0.088 & 0.171 & 0.111 & 0.100 \\
CNN-LSTM & 0.138 & 0.105 & 0.160 & 0.125 & 0.182 & 0.142 & 0.188 & 0.126 & 0.136 & 0.171 & 0.155 & 0.155 & 0.169 & 0.157 & 0.212 & 0.180 & 0.154 & 0.174 & 0.157 \\
SSPGKAN & 0.160 & 0.193 & 0.207 & 0.288 & 0.165 & 0.203 & 0.283 & 0.156 & 0.165 & 0.174 & 0.177 & 0.191 & 0.283 & 0.286 & 0.262 & 0.274 & 0.288 & 0.279 & 0.224 \\
Ours & \textbf{0.013} & \textbf{0.027} & \textbf{0.019} & \textbf{0.016} & \textbf{0.049} & \textbf{0.025} & 0.171 & \textbf{0.037} & \textbf{0.054} & \textbf{0.045} & \textbf{0.035} & \textbf{0.068} & 0.216 & 0.125 & \textbf{0.032} & 0.109 & \textbf{0.088} & 0.114 & \textbf{0.069} \\
\hline
\end{tabular}
}
\end{center}

\begin{center}
    \includegraphics[width=0.95\textwidth]{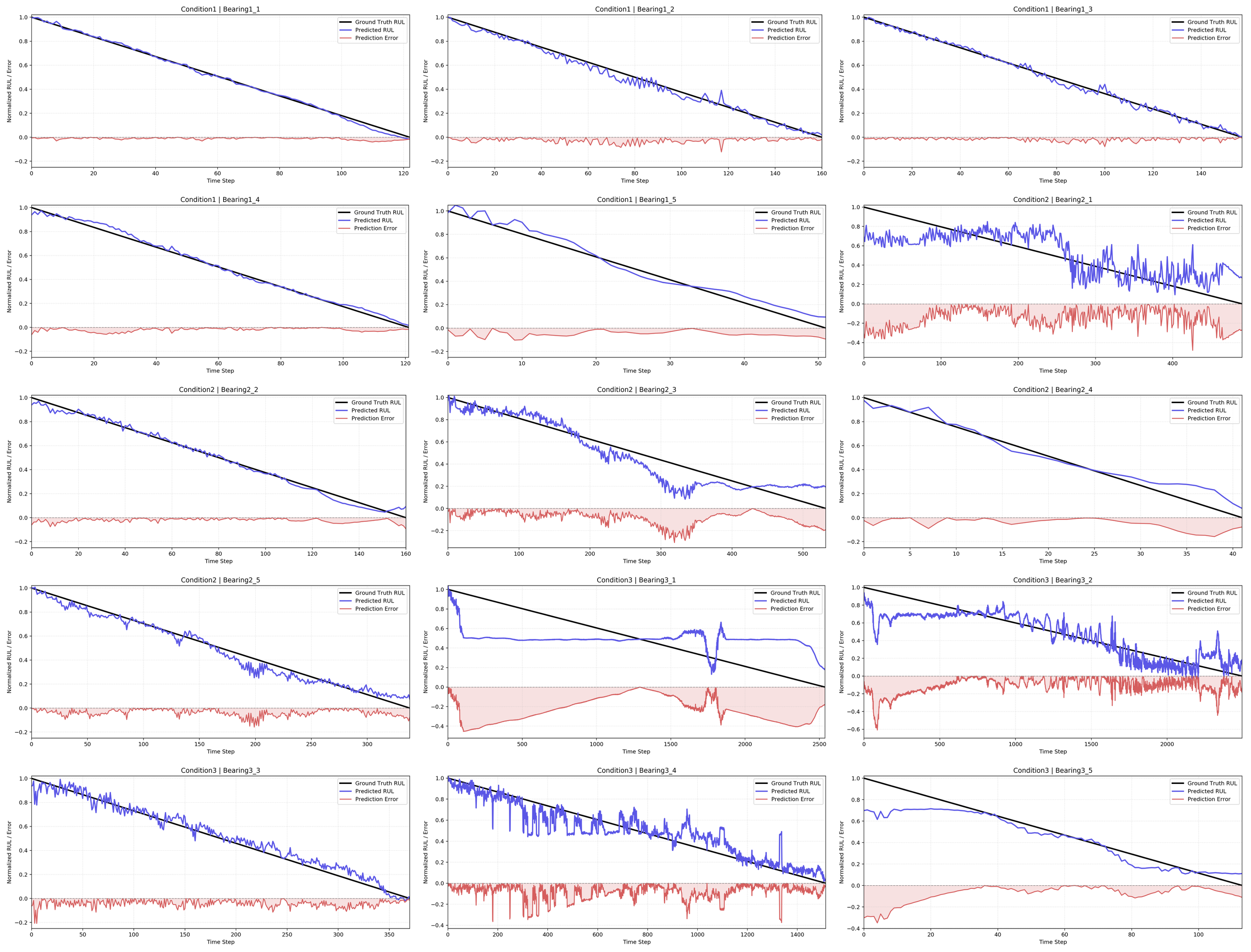}
    \captionof{figure}{RUL prediction results for test bearings in the XJTU-SY dataset.}
    \label{fig:xjtu_sy_rul}
\end{center}

\begin{center}
\captionof{table}{Comparison results of MAE on the XJTU-SY dataset.}
\label{tab:mae_xjtu}
\scriptsize
\resizebox{\textwidth}{!}{
\begin{tabular}{lccccccccccccccccccc}
\hline
Method & Br1\_1 & Br1\_2 & Br1\_3 & Br1\_4 & Br1\_5 & C1\_Avg & Br2\_1 & Br2\_2 & Br2\_3 & Br2\_4 & Br2\_5 & C2\_Avg & Br3\_1 & Br3\_2 & Br3\_3 & Br3\_4 & Br3\_5 & C3\_Avg & Total\_Avg \\
\hline
T-GCN & 0.098 & 0.081 & 0.121 & 0.116 & 0.137 & 0.111 & 0.107 & 0.099 & 0.096 & 0.108 & 0.119 & 0.106 & 0.128 & 0.153 & 0.127 & 0.100 & 0.173 & 0.136 & 0.118 \\
ChebGCN-LSTM & 0.091 & 0.128 & 0.139 & 0.145 & 0.151 & 0.131 & 0.104 & 0.095 & 0.133 & 0.137 & 0.108 & 0.115 & 0.139 & 0.123 & 0.087 & 0.096 & 0.167 & 0.122 & 0.123 \\
SAGCN-SA & 0.087 & 0.092 & 0.099 & 0.073 & 0.138 & 0.098 & 0.110 & 0.094 & 0.102 & 0.107 & 0.085 & 0.100 & 0.139 & 0.142 & 0.090 & 0.099 & 0.160 & 0.126 & 0.108 \\
MR-LSTM & 0.076 & 0.086 & 0.115 & 0.062 & 0.096 & 0.087 & 0.074 & 0.083 & 0.077 & 0.075 & 0.094 & 0.081 & 0.124 & 0.135 & 0.080 & 0.076 & 0.158 & 0.115 & 0.094 \\
SDFEPN & 0.073 & 0.079 & 0.094 & 0.081 & 0.123 & 0.090 & 0.093 & 0.066 & 0.104 & 0.067 & 0.085 & 0.083 & \textbf{0.088} & 0.124 & 0.054 & 0.075 & 0.143 & 0.097 & 0.090 \\
CBAM-MSCNN & 0.075 & 0.082 & 0.111 & 0.084 & 0.128 & 0.096 & \textbf{0.071} & 0.079 & 0.107 & 0.074 & 0.093 & 0.085 & 0.119 & \textbf{0.089} & 0.077 & \textbf{0.044} & 0.134 & 0.093 & 0.091 \\
DyWave-BiAGCN & 0.068 & 0.061 & 0.100 & 0.067 & 0.098 & 0.079 & 0.080 & 0.071 & 0.084 & 0.058 & 0.066 & 0.072 & \textbf{0.088} & 0.091 & 0.062 & 0.048 & 0.127 & \textbf{0.083} & 0.078 \\
DC-DGCN & 0.084 & 0.055 & 0.067 & 0.094 & 0.124 & 0.085 & 0.085 & 0.064 & 0.106 & 0.059 & 0.075 & 0.078 & 0.091 & 0.117 & 0.109 & 0.089 & 0.140 & 0.109 & 0.091 \\
GCN & 0.121 & 0.101 & 0.119 & 0.105 & 0.152 & 0.120 & 0.119 & 0.124 & 0.121 & 0.127 & 0.107 & 0.120 & 0.153 & 0.188 & 0.152 & 0.135 & 0.196 & 0.165 & 0.135 \\
STACP-GCN & 0.064 & 0.050 & 0.059 & 0.081 & 0.112 & 0.073 & 0.064 & 0.072 & 0.084 & 0.081 & 0.057 & 0.072 & 0.100 & 0.077 & 0.072 & 0.072 & 0.126 & 0.089 & 0.078 \\
CNN-LSTM & 0.124 & 0.090 & 0.135 & 0.108 & 0.157 & 0.123 & 0.158 & 0.105 & 0.112 & 0.134 & 0.126 & 0.127 & 0.119 & 0.130 & 0.177 & 0.147 & 0.119 & 0.138 & 0.129 \\
SSPGKAN & 0.132 & 0.169 & 0.166 & 0.247 & 0.138 & 0.170 & 0.245 & 0.131 & 0.140 & 0.139 & 0.141 & 0.159 & 0.248 & 0.247 & 0.224 & 0.236 & 0.253 & 0.242 & 0.190 \\
Ours & \textbf{0.010} & \textbf{0.021} & \textbf{0.014} & \textbf{0.012} & \textbf{0.042} & \textbf{0.020} & 0.142 & \textbf{0.031} & \textbf{0.044} & \textbf{0.037} & \textbf{0.029} & \textbf{0.057} & 0.173 & 0.091 & \textbf{0.026} & 0.082 & \textbf{0.076} & 0.099 & \textbf{0.055} \\
\hline
\end{tabular}
}
\end{center}

\begin{center}
\captionof{table}{Comparison results of EAS on the XJTU-SY dataset.}
\label{tab:eas_xjtu}
\scriptsize
\resizebox{\textwidth}{!}{
\begin{tabular}{lccccccccccccccccccc}
\hline
Method & Br1\_1 & Br1\_2 & Br1\_3 & Br1\_4 & Br1\_5 & C1\_Avg & Br2\_1 & Br2\_2 & Br2\_3 & Br2\_4 & Br2\_5 & C2\_Avg & Br3\_1 & Br3\_2 & Br3\_3 & Br3\_4 & Br3\_5 & C3\_Avg & Total\_Avg \\
\hline
T-GCN & 0.0113 & 0.0103 & 0.0147 & 0.0128 & 0.0138 & 0.0126 & 0.0133 & 0.0164 & 0.0132 & 0.0112 & 0.0132 & 0.0134 & 0.0095 & 0.0122 & 0.0114 & 0.0145 & 0.0122 & 0.0119 & 0.0126 \\
ChebGCN-LSTM & 0.0126 & 0.0108 & 0.0164 & 0.0133 & 0.0142 & 0.0134 & 0.0113 & 0.0154 & 0.0108 & 0.0115 & 0.0141 & 0.0126 & 0.0109 & 0.0073 & 0.0124 & 0.0131 & 0.0129 & 0.0113 & 0.0125 \\
SAGCN-SA & 0.0111 & 0.0082 & 0.0131 & 0.0093 & 0.0123 & 0.0108 & 0.0097 & 0.0134 & 0.0127 & 0.0089 & 0.0108 & 0.0111 & 0.0099 & 0.0110 & 0.0131 & 0.0145 & 0.0117 & 0.0120 & 0.0113 \\
MR-LSTM & 0.0080 & 0.0082 & 0.0105 & 0.0085 & 0.0111 & 0.0092 & 0.0067 & 0.0105 & 0.0074 & 0.0087 & 0.0094 & 0.0085 & 0.0077 & 0.0069 & 0.0111 & 0.0115 & 0.0099 & 0.0094 & 0.0090 \\
SDFEPN & 0.0071 & 0.0077 & 0.0090 & 0.0084 & 0.0112 & 0.0086 & 0.0085 & 0.0101 & 0.0093 & 0.0082 & 0.0084 & 0.0089 & 0.0069 & 0.0075 & 0.0111 & 0.0120 & 0.0091 & 0.0093 & 0.0089 \\
CBAM-MSCNN & 0.0085 & 0.0083 & 0.0107 & 0.0094 & 0.0123 & 0.0098 & \textbf{0.0057} & 0.0105 & 0.0111 & 0.0082 & 0.0088 & 0.0088 & 0.0077 & 0.0079 & 0.0115 & 0.0119 & 0.0102 & 0.0098 & 0.0095 \\
DyWave-BiAGCN & 0.0055 & 0.0059 & 0.0089 & 0.0067 & 0.0096 & 0.0073 & 0.0074 & 0.0085 & 0.0081 & 0.0063 & 0.0070 & 0.0074 & \textbf{0.0048} & \textbf{0.0052} & 0.0093 & 0.0100 & 0.0078 & \textbf{0.0074} & 0.0074 \\
DC-DGCN & 0.0136 & 0.0086 & 0.0109 & 0.0137 & 0.0115 & 0.0117 & 0.0115 & 0.0103 & 0.0083 & 0.0133 & 0.0102 & 0.0107 & 0.0083 & 0.0163 & 0.0082 & 0.0083 & 0.0223 & 0.0127 & 0.0117 \\
GCN & 0.0156 & 0.0157 & 0.0162 & 0.0127 & 0.0139 & 0.0148 & 0.0179 & 0.0133 & 0.0158 & 0.0183 & 0.0136 & 0.0158 & 0.0193 & 0.0222 & 0.0136 & 0.0220 & 0.0222 & 0.0199 & 0.0168 \\
STACP-GCN & 0.0109 & 0.0088 & 0.0103 & 0.0092 & 0.0072 & 0.0093 & 0.0114 & 0.0077 & 0.0098 & 0.0119 & 0.0083 & 0.0098 & 0.0217 & 0.0123 & 0.0067 & 0.0227 & 0.0194 & 0.0166 & 0.0119 \\
CNN-LSTM & 0.0162 & 0.0132 & 0.0158 & 0.0159 & 0.0121 & 0.0146 & 0.0168 & 0.0131 & 0.0141 & 0.0174 & 0.0138 & 0.0150 & 0.0175 & 0.0224 & 0.0115 & 0.0225 & 0.0223 & 0.0192 & 0.0163 \\
SSPGKAN & 0.0120 & 0.0142 & 0.0154 & 0.0220 & 0.0123 & 0.0152 & 0.0215 & 0.0120 & 0.0127 & 0.0128 & 0.0112 & 0.0140 & 0.0225 & 0.0223 & 0.0202 & 0.0214 & 0.0215 & 0.0216 & 0.0169 \\
Ours & \textbf{0.0008} & \textbf{0.0019} & \textbf{0.0012} & \textbf{0.0011} & \textbf{0.0042} & \textbf{0.0019} & 0.0124 & \textbf{0.0025} & \textbf{0.0038} & \textbf{0.0037} & \textbf{0.0025} & \textbf{0.0050} & 0.0158 & 0.0075 & \textbf{0.0023} & \textbf{0.0073} & \textbf{0.0065} & 0.0079 & \textbf{0.0049} \\
\hline
\end{tabular}
}
\end{center}

Table~\ref{tab:mae_xjtu} further evaluates the average absolute prediction deviation. PE-BMGN obtains the lowest Total\_Avg MAE of 0.055, which is lower than the best baseline value of 0.078 achieved by DyWave-BiAGCN and STACP-GCN, and substantially lower than SSPGKAN with 0.190. Under Condition I, PE-BMGN achieves a C1\_Avg MAE of 0.020 and obtains the best result on all five test bearings, showing that the proposed framework can provide highly stable prediction when the degradation patterns are relatively coherent. Under Condition II, the proposed method also obtains the lowest C2\_Avg MAE of 0.057, outperforming DyWave-BiAGCN with 0.072 and SSPGKAN with 0.159. This result verifies the effectiveness of the memory-augmented prediction network, because the final prediction stage does not rely only on the current latent representation, but retrieves relevant historical degradation prototypes to provide additional matching information. Under Condition III, PE-BMGN obtains a C3\_Avg MAE of 0.099, which is higher than DyWave-BiAGCN with 0.083. This indicates that the proposed memory-enhanced prediction module improves average prediction stability in most cases, but the benefit of memory retrieval may be weakened when the test trajectory contains strong distribution shift or insufficiently matched historical degradation prototypes.

Table~\ref{tab:eas_xjtu} shows the comparison results in terms of EAS, which reflects not only numerical errors but also the practical engineering risk caused by early or late RUL prediction. PE-BMGN achieves the lowest Total\_Avg EAS of 0.0049, outperforming DyWave-BiAGCN with 0.0074 and SSPGKAN with 0.0169. The proposed method obtains the best C1\_Avg and C2\_Avg EAS values of 0.0019 and 0.0050, respectively, which is consistent with the RMSE and MAE results. This consistency indicates that PE-BMGN does not merely reduce point-wise regression errors, but also produces more reliable degradation-to-RUL mappings from the perspective of engineering-risk-sensitive evaluation. Compared with SSPGKAN, the clear EAS advantage further demonstrates that the proposed KAN usage is more effective: instead of using KAN mainly for degradation stage classification and sparse path graph transmission, PE-BMGN jointly integrates KAN-based nonlinear mapping into bidirectional graph dependency extraction, memory-enhanced prediction, and physics-enhanced loss. The regularization term constrains the spline coefficient magnitude and coefficient distribution of KAN layers, which helps suppress redundant functional responses and encourages compact nonlinear mappings. Nevertheless, the C3\_Avg EAS of PE-BMGN is 0.0079, slightly higher than DyWave-BiAGCN with 0.0074, showing that irregular trajectories such as Br3\_1 remain challenging for engineering-risk-sensitive prediction.

Figure \ref{fig:xjtu_sy_rul} provides trajectory-level evidence for the above conclusions. For most bearings, the predicted RUL curves generally follow the degradation tendency of the ground truth, which explains why PE-BMGN achieves the best overall results across RMSE, MAE, and EAS. This trajectory consistency is closely related to the design of the proposed framework: the bidirectional multi-order graph fusion network models forward and backward degradation dependencies through functional mappings, the memory-augmented prediction network introduces historical degradation prototypes, and the KAN-based regressor performs the final nonlinear RUL mapping. Meanwhile, several curves still show visible deviations, which correspond to the larger errors observed for Br2\_1 and Br3\_1 in the tables. Thus, Figure \ref{fig:xjtu_sy_rul} confirms that the proposed method captures the global degradation trend well, while also revealing the remaining difficulty of handling abrupt or atypical degradation evolution.

\subsubsection{Interpretability experiments}

To further explain the prediction behavior of the proposed PE-BMGN on the XJTU-SY dataset, two representative bearings are selected for functional response analysis. Br1\_3 is selected as a representative favorable case, while Br2\_1 is selected as a representative challenging case. Although Br1\_1 and Br1\_4 obtain slightly lower numerical errors than Br1\_3, the learned functions of Br1\_3 exhibit clearer cross-module consistency. Similarly, although Br3\_1 has the largest numerical error, Br2\_1 provides more typical functional patterns for explaining difficult prediction behavior.

\begin{center}
    \includegraphics[width=0.95\textwidth]{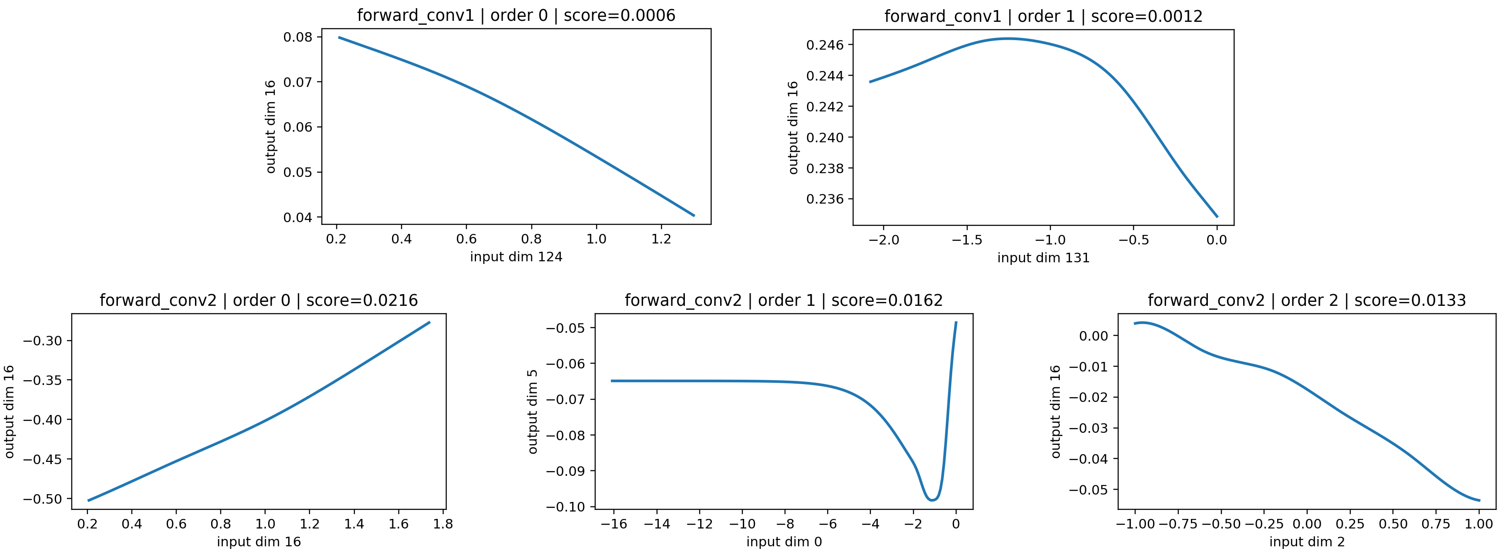}\par
    {\small (a) Forward graph propagation.\par}
    \vspace{0.2em}
    \includegraphics[width=0.95\textwidth]{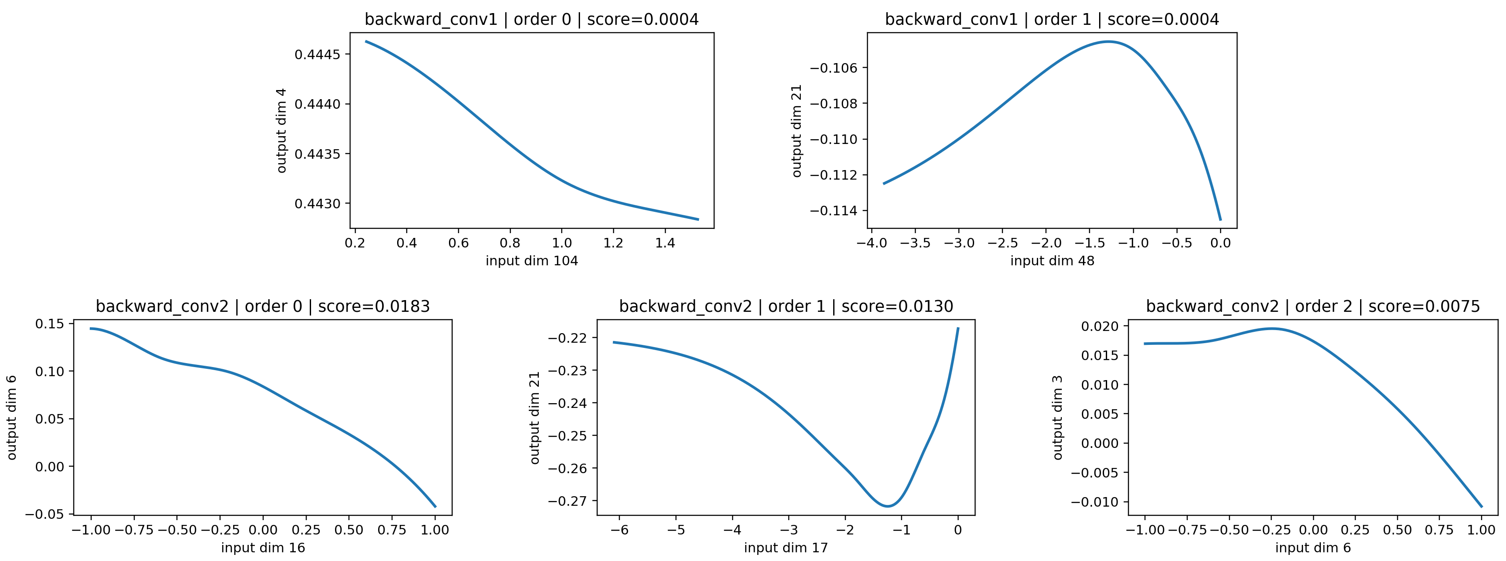}\par
    {\small (b) Backward graph propagation.\par}
    \vspace{0.2em}
    \includegraphics[width=0.7\textwidth]{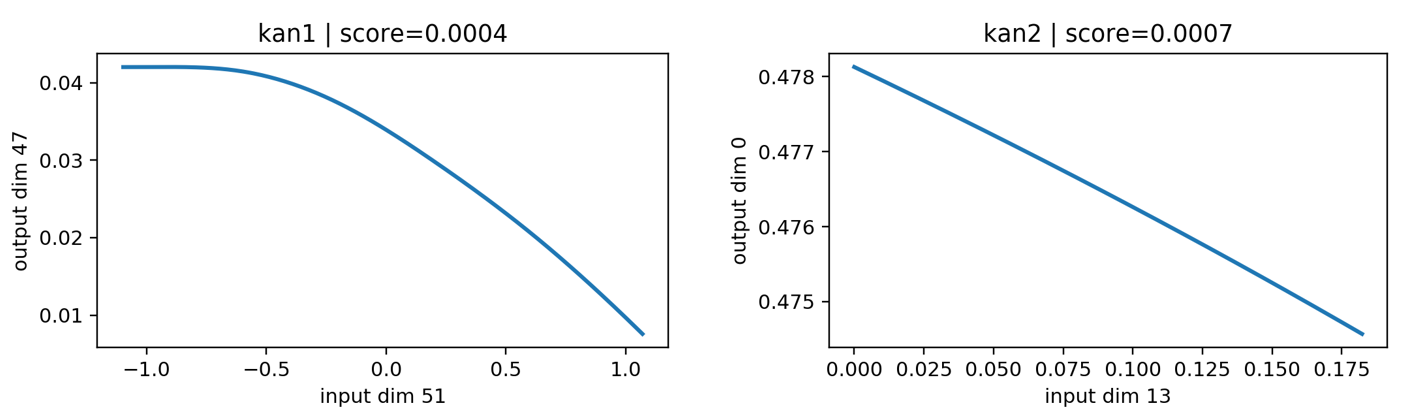}\par
    {\small (c) KAN regressor.\par}
    \captionof{figure}{Learned function responses of Br1\_3 (favorable case) on the XJTU-SY dataset. All mappings are smooth, which shows that our method can model nonlinear degradation processes in a stable manner.}
    \label{fig:xjtu_br13_responses}
\end{center}

For Br1\_3, the prediction errors are relatively low, with RMSE, MAE, and EAS values of 0.019, 0.014, and 0.0012, respectively. As shown in Figure \ref{fig:xjtu_br13_responses}, the learned functions in both the forward and backward graph propagation branches show continuous and stable response patterns. Most mappings are monotonic or weakly nonlinear, and no obvious abrupt oscillations or boundary jumps can be observed. This indicates that the bidirectional multi-order graph fusion network extracts degradation dependencies in a stable manner. In addition, the responses from shallower to deeper transformations show a reasonable hierarchy: the shallower mappings mainly preserve smooth basic trends, while the deeper mappings introduce more active but still continuous nonlinear responses. This phenomenon suggests that the model does not rely on unstable local fluctuations to fit Br1\_3, but gradually refines the degradation representation through functional graph propagation.

\begin{center}
    \includegraphics[width=0.95\textwidth]{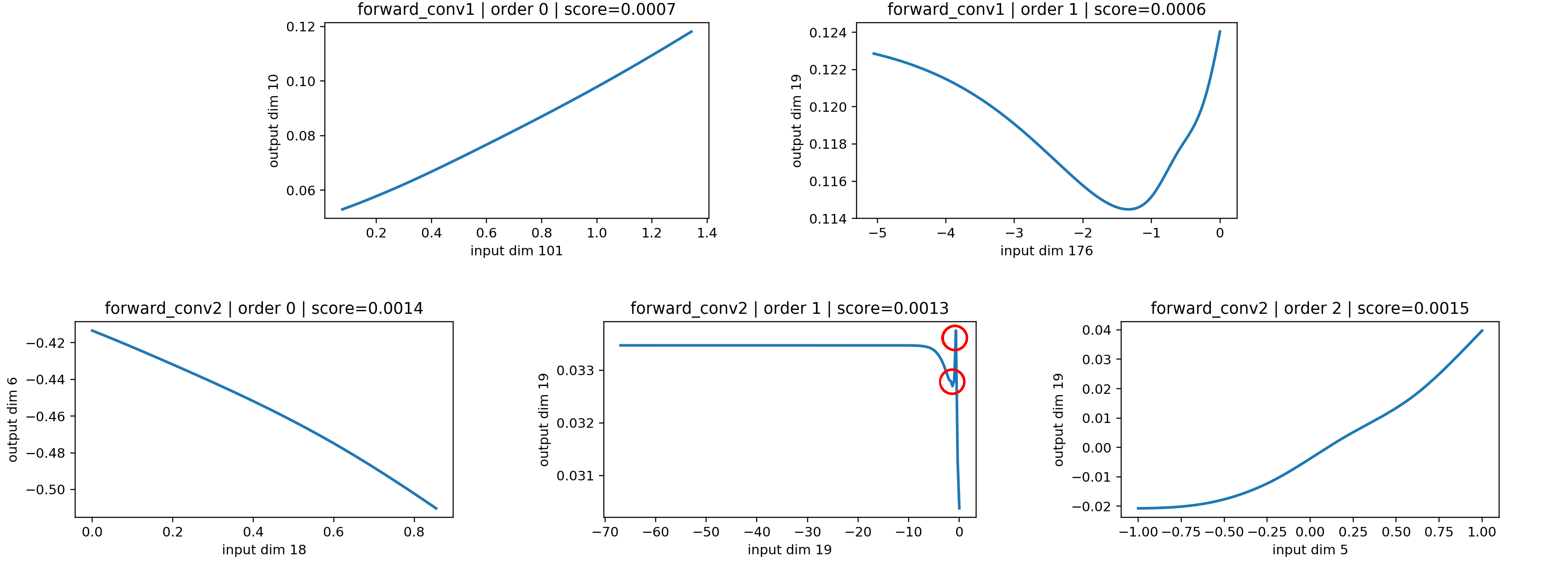}\par
    {\small (a) Forward graph propagation.\par}
    \vspace{0.2em}
    \includegraphics[width=0.95\textwidth]{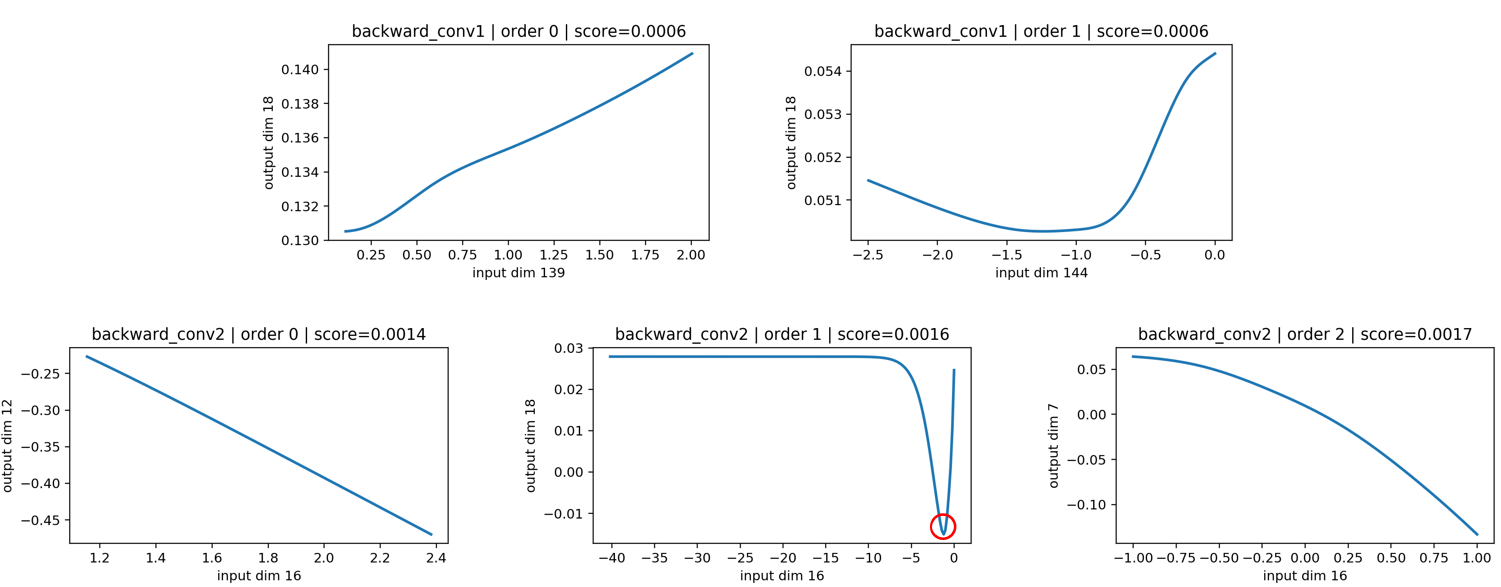}\par
    {\small (b) Backward graph propagation.\par}
    \vspace{0.2em}
    \includegraphics[width=0.7\textwidth]{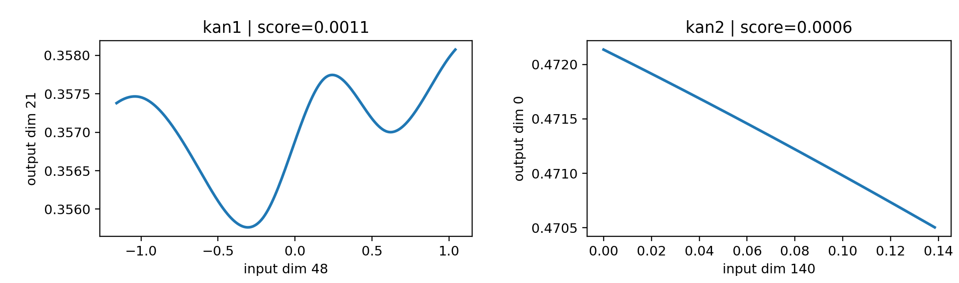}\par
    {\small (c) KAN regressor.\par}
    \captionof{figure}{Learned function responses of Br2\_1 (challenge case) on the XJTU-SY dataset. The sharp changes within the red circle show that the model has captured significant local mutations or boundary-sensitive degradation features in this input interval, suggesting that the degradation process of this bearing is more unstable and therefore more difficult to predict.}
    \label{fig:xjtu_br21_responses}
\end{center}

The final KAN regressor of Br1\_3 further supports this interpretation. Its learned functions are also smooth and continuous, indicating that the final prediction stage only needs to perform a stable nonlinear mapping from the fused degradation representation to the RUL output. Therefore, the favorable prediction performance of Br1\_3 can be attributed to a coherent functional chain: the bidirectional graph propagation branches extract stable degradation features, and the KAN regressor converts these features into RUL values through smooth functional mappings. This consistency among graph-level representation learning and output regression provides an interpretable explanation for why the proposed method can achieve accurate prediction on this bearing.

For Br2\_1, the prediction errors are much larger, with RMSE, MAE, and EAS values of 0.171, 0.142, and 0.0124, respectively. As shown in Figure \ref{fig:xjtu_br21_responses}, the learned functions in the bidirectional graph propagation branches present more complex local variations than those of Br1\_3. Several mappings, especially in the deeper graph transformations, show stronger high-order responses, endpoint-sensitive changes, and locally concentrated peaks. These patterns indicate that the degradation trajectory of Br2\_1 is more difficult to describe by smooth and globally consistent functional mappings. The fact that similar unstable responses appear in both forward and backward branches also suggests that the difficulty is not caused by a single propagation direction, but is related to the intrinsic irregularity of the degradation representation learned from this bearing.

Compared with Br1\_3, the final KAN regressor of Br2\_1 remains relatively smooth, but it cannot fully compensate for the less stable graph-level responses produced by the bidirectional graph propagation branches. This means that the main source of prediction difficulty lies before the final regression stage: the forward and backward branches have already encoded more boundary-sensitive and locally fluctuating degradation features. Once such features are fused and passed to the KAN regressor, the output mapping can only smooth the final transformation to a limited extent, but cannot completely recover a coherent degradation trajectory. Therefore, the larger prediction deviation of Br2\_1 can be explained by the weaker consistency between graph-level functional representation and final RUL regression.

The comparison between Br1\_3 and Br2\_1 demonstrates that the interpretability of PE-BMGN comes from observing the entire functional prediction chain rather than a single learned function. When the forward graph propagation branch, backward graph propagation branch, and KAN regressor produce coherent and smooth mappings, the model tends to generate accurate and stable RUL predictions. In contrast, when the graph-level functions contain strong local peaks, endpoint sensitivity, or inconsistent nonlinear responses, the final KAN regressor has limited ability to eliminate the accumulated representation uncertainty. These results verify that the proposed method not only improve prediction accuracy, but also provide a functional perspective for explaining why the model performs well on some bearings and less favorably on more challenging degradation trajectories.

\subsection{Case study 2: PHM2012 bearing dataset}

\subsubsection{Comparative experiments}

As shown in Table~\ref{tab:rmse_phm2012}, the proposed PE-BMGN obtains the lowest Total\_Avg RMSE of 0.095 on the PHM2012 dataset, outperforming all baselines. This indicates that the proposed model reduces large-deviation prediction errors under the original PHM2012 challenge setting. The advantage is particularly evident under Condition II and Condition III. Under Condition II, PE-BMGN achieves a C2\_Avg RMSE of 0.078, which is clearly lower than the best baseline value of 0.129 obtained by STACP-GCN. For Br3\_3, the proposed method also obtains the lowest RMSE of 0.052, showing strong generalization under Condition III. Under Condition I, the improvement is more moderate, with a C1\_Avg RMSE of 0.120 compared with 0.123 achieved by DyWave-BiAGCN. This result suggests that the bidirectional multi-order graph fusion network can effectively reduce large errors when the degradation dependency can be represented by coherent nonlinear graph functions, while Condition I still contains several trajectories that are more difficult to model.

\begin{center}
\captionof{table}{Comparison results of RMSE on the PHM2012 dataset.}
\label{tab:rmse_phm2012}
\scriptsize
\resizebox{\textwidth}{!}{
\begin{tabular}{lcccccccccccccc}
\hline
Method & Br1\_3 & Br1\_4 & Br1\_5 & Br1\_6 & Br1\_7 & C1\_Avg & Br2\_3 & Br2\_4 & Br2\_5 & Br2\_6 & Br2\_7 & C2\_Avg & Br3\_3 & Total\_Avg \\
\hline
T-GCN & 0.140 & 0.158 & 0.171 & 0.179 & 0.173 & 0.164 & 0.123 & 0.122 & 0.205 & 0.144 & 0.250 & 0.169 & 0.121 & 0.162 \\
ChebGCN-LSTM & 0.133 & 0.145 & 0.189 & 0.178 & 0.162 & 0.161 & 0.115 & 0.132 & 0.194 & 0.172 & 0.268 & 0.176 & 0.142 & 0.166 \\
SAGCN-SA & 0.121 & 0.135 & 0.173 & 0.165 & 0.151 & 0.149 & 0.105 & 0.125 & 0.182 & 0.160 & 0.239 & 0.162 & 0.131 & 0.153 \\
MR-LSTM & \textbf{0.092} & 0.116 & 0.162 & 0.148 & 0.134 & 0.130 & \textbf{0.080} & 0.117 & 0.156 & 0.133 & 0.230 & 0.143 & 0.114 & 0.135 \\
SDFEPN & 0.105 & 0.107 & 0.156 & 0.155 & 0.143 & 0.133 & 0.102 & 0.096 & 0.168 & 0.147 & 0.232 & 0.149 & 0.130 & 0.140 \\
CBAM-MSCNN & 0.114 & 0.126 & 0.150 & \textbf{0.131} & 0.138 & 0.132 & 0.092 & 0.129 & 0.174 & 0.117 & 0.238 & 0.150 & 0.100 & 0.137 \\
DyWave-BiAGCN & 0.097 & 0.109 & \textbf{0.144} & 0.138 & 0.125 & 0.123 & 0.082 & 0.101 & 0.150 & 0.133 & 0.197 & 0.133 & 0.104 & 0.125 \\
DC-DGCN & 0.110 & 0.112 & 0.161 & 0.160 & 0.148 & 0.138 & 0.107 & 0.101 & 0.173 & 0.152 & 0.239 & 0.154 & 0.135 & 0.145 \\
GCN & 0.143 & 0.157 & 0.170 & 0.178 & 0.175 & 0.165 & 0.126 & 0.125 & 0.207 & 0.145 & 0.253 & 0.171 & 0.124 & 0.164 \\
STACP-GCN & 0.102 & 0.113 & 0.149 & 0.143 & 0.130 & 0.127 & 0.087 & 0.106 & 0.145 & 0.124 & 0.183 & 0.129 & 0.109 & 0.126 \\
CNN-LSTM & 0.137 & 0.148 & 0.193 & 0.183 & 0.167 & 0.166 & 0.120 & 0.135 & 0.199 & 0.177 & 0.273 & 0.181 & 0.147 & 0.171 \\
SSPGKAN & 0.177 & 0.229 & 0.306 & 0.303 & 0.281 & 0.259 & 0.324 & 0.296 & 0.311 & 0.298 & 0.299 & 0.306 & 0.303 & 0.284 \\
Ours & \textbf{0.092} & \textbf{0.099} & 0.159 & 0.142 & \textbf{0.107} & \textbf{0.120} & 0.085 & \textbf{0.072} & \textbf{0.128} & \textbf{0.054} & \textbf{0.052} & \textbf{0.078} & \textbf{0.052} & \textbf{0.095} \\
\hline
\end{tabular}
}
\end{center}

Table~\ref{tab:mae_phm2012} further verifies the average prediction stability of PE-BMGN. The proposed method achieves the lowest Total\_Avg MAE of 0.074, reducing the value from 0.106 obtained by DyWave-BiAGCN. This improvement is consistent with the design motivation of the memory-augmented prediction network, which introduces historical degradation prototypes into the final regression stage instead of relying only on the current latent representation. At the condition level, PE-BMGN obtains the lowest MAE values under all three operating conditions, with C1\_Avg, C2\_Avg, and Br3\_3 values of 0.094, 0.060, and 0.042, respectively. Especially under Condition II, the proposed method achieves the best result on all five test bearings. This demonstrates that the combination of memory retrieval and KAN-based regression improves the stability of RUL estimation for most degradation trajectories.

\begin{center}
\captionof{table}{Comparison results of MAE on the PHM2012 dataset.}
\label{tab:mae_phm2012}
\scriptsize
\resizebox{\textwidth}{!}{
\begin{tabular}{lcccccccccccccc}
\hline
Method & Br1\_3 & Br1\_4 & Br1\_5 & Br1\_6 & Br1\_7 & C1\_Avg & Br2\_3 & Br2\_4 & Br2\_5 & Br2\_6 & Br2\_7 & C2\_Avg & Br3\_3 & Total\_Avg \\
\hline
T-GCN & 0.121 & 0.134 & 0.153 & 0.163 & 0.149 & 0.144 & 0.102 & 0.102 & 0.172 & 0.125 & 0.218 & 0.144 & 0.100 & 0.140 \\
ChebGCN-LSTM & 0.113 & 0.123 & 0.159 & 0.151 & 0.138 & 0.137 & 0.095 & 0.110 & 0.163 & 0.145 & 0.224 & 0.147 & 0.120 & 0.140 \\
SAGCN-SA & 0.103 & 0.115 & 0.147 & 0.139 & 0.128 & 0.126 & 0.087 & 0.105 & 0.154 & 0.135 & 0.200 & 0.136 & 0.111 & 0.129 \\
MR-LSTM & 0.078 & 0.097 & 0.142 & 0.123 & 0.110 & 0.110 & 0.068 & 0.099 & 0.132 & 0.115 & 0.198 & 0.122 & 0.097 & 0.114 \\
SDFEPN & 0.088 & 0.089 & 0.132 & 0.127 & 0.117 & 0.111 & 0.086 & 0.080 & 0.139 & 0.121 & 0.191 & 0.123 & 0.115 & 0.117 \\
CBAM-MSCNN & 0.096 & 0.105 & 0.127 & 0.111 & 0.115 & 0.111 & 0.077 & 0.104 & 0.150 & 0.099 & 0.204 & 0.127 & 0.086 & 0.116 \\
DyWave-BiAGCN & 0.081 & 0.090 & \textbf{0.121} & 0.118 & 0.104 & 0.103 & 0.067 & 0.085 & 0.123 & 0.117 & 0.176 & 0.114 & 0.086 & 0.106 \\
DC-DGCN & 0.093 & 0.096 & 0.134 & 0.140 & 0.122 & 0.117 & 0.088 & 0.083 & 0.150 & 0.126 & 0.197 & 0.129 & 0.108 & 0.122 \\
GCN & 0.122 & 0.131 & 0.145 & 0.152 & 0.151 & 0.140 & 0.105 & 0.105 & 0.171 & 0.119 & 0.203 & 0.141 & 0.104 & 0.137 \\
STACP-GCN & 0.085 & 0.095 & 0.125 & 0.123 & 0.106 & 0.107 & 0.071 & 0.089 & 0.120 & 0.106 & 0.160 & 0.109 & 0.092 & 0.107 \\
CNN-LSTM & 0.115 & 0.121 & 0.162 & 0.155 & 0.142 & 0.139 & 0.102 & 0.114 & 0.172 & 0.150 & 0.225 & 0.153 & 0.124 & 0.144 \\
SSPGKAN & 0.132 & 0.192 & 0.263 & 0.259 & 0.238 & 0.217 & 0.271 & 0.258 & 0.267 & 0.260 & 0.257 & 0.263 & 0.261 & 0.242 \\
Ours & \textbf{0.071} & \textbf{0.082} & 0.124 & \textbf{0.106} & \textbf{0.086} & \textbf{0.094} & \textbf{0.062} & \textbf{0.052} & \textbf{0.097} & \textbf{0.044} & \textbf{0.045} & \textbf{0.060} & \textbf{0.042} & \textbf{0.074} \\
\hline
\end{tabular}
}
\end{center}

The EAS comparison in Table~\ref{tab:eas_phm2012} shows that PE-BMGN also achieves the best engineering-oriented prediction performance. The Total\_Avg EAS of the proposed method is 0.0067, which is lower than 0.0093 of STACP-GCN and 0.0094 of DyWave-BiAGCN. Since EAS is more sensitive to the engineering risk caused by biased RUL prediction, this result indicates that PE-BMGN not only decreases numerical prediction errors but also improves the reliability of maintenance-oriented prediction. The advantage is especially clear under Condition II, where the C2\_Avg EAS decreases to 0.0054, and under Condition III, where Br3\_3 obtains the lowest EAS of 0.0035. However, under Condition I, the proposed method does not achieve the best EAS on every individual bearing, such as Br1\_5 and Br1\_6. Therefore, the EAS results show that the proposed method improves overall engineering reliability, but local degradation fluctuations still affect several difficult trajectories.

\begin{center}
\captionof{table}{Comparison results of EAS on the PHM2012 dataset.}
\label{tab:eas_phm2012}
\scriptsize
\resizebox{\textwidth}{!}{
\begin{tabular}{lcccccccccccccc}
\hline
Method & Br1\_3 & Br1\_4 & Br1\_5 & Br1\_6 & Br1\_7 & C1\_Avg & Br2\_3 & Br2\_4 & Br2\_5 & Br2\_6 & Br2\_7 & C2\_Avg & Br3\_3 & Total\_Avg \\
\hline
T-GCN & 0.0166 & 0.0169 & 0.0124 & 0.0126 & 0.0120 & 0.0141 & 0.0157 & 0.0136 & 0.0157 & 0.0129 & 0.0159 & 0.0147 & 0.0149 & 0.0145 \\
ChebGCN-LSTM & 0.0135 & 0.0161 & 0.0107 & 0.0120 & 0.0158 & 0.0136 & 0.0132 & 0.0132 & 0.0158 & 0.0110 & 0.0164 & 0.0139 & 0.0168 & 0.0140 \\
SAGCN-SA & 0.0160 & 0.0151 & 0.0110 & 0.0117 & 0.0127 & 0.0133 & 0.0139 & 0.0118 & 0.0142 & 0.0117 & 0.0121 & 0.0127 & 0.0161 & 0.0133 \\
MR-LSTM & 0.0102 & 0.0141 & 0.0096 & 0.0078 & 0.0104 & 0.0104 & 0.0108 & 0.0107 & 0.0114 & 0.0096 & 0.0102 & 0.0105 & 0.0132 & 0.0107 \\
SDFEPN & 0.0138 & 0.0145 & 0.0099 & 0.0093 & 0.0106 & 0.0116 & 0.0128 & 0.0090 & 0.0126 & 0.0103 & 0.0119 & 0.0113 & 0.0137 & 0.0116 \\
CBAM-MSCNN & 0.0119 & 0.0125 & 0.0084 & \textbf{0.0060} & 0.0104 & 0.0098 & 0.0105 & 0.0093 & 0.0124 & 0.0062 & 0.0113 & 0.0099 & 0.0133 & 0.0102 \\
DyWave-BiAGCN & 0.0110 & 0.0120 & \textbf{0.0073} & 0.0068 & 0.0089 & 0.0092 & 0.0099 & 0.0084 & 0.0104 & 0.0078 & 0.0094 & 0.0092 & 0.0115 & 0.0094 \\
DC-DGCN & 0.0136 & 0.0086 & 0.0109 & 0.0137 & 0.0115 & 0.0117 & 0.0115 & 0.0103 & 0.0083 & 0.0133 & 0.0102 & 0.0107 & 0.0082 & 0.0109 \\
GCN & 0.0156 & 0.0157 & 0.0162 & 0.0127 & 0.0139 & 0.0148 & 0.0179 & 0.0133 & 0.0158 & 0.0183 & 0.0136 & 0.0158 & 0.0136 & 0.0151 \\
STACP-GCN & 0.0109 & 0.0088 & 0.0103 & 0.0092 & \textbf{0.0072} & 0.0093 & 0.0114 & 0.0077 & 0.0098 & 0.0119 & 0.0083 & 0.0098 & 0.0067 & 0.0093 \\
CNN-LSTM & 0.0162 & 0.0132 & 0.0158 & 0.0159 & 0.0121 & 0.0146 & 0.0168 & 0.0131 & 0.0141 & 0.0174 & 0.0138 & 0.0150 & 0.0115 & 0.0145 \\
SSPGKAN & 0.0111 & 0.0166 & 0.0247 & 0.0241 & 0.0221 & 0.0197 & 0.0246 & 0.0233 & 0.0241 & 0.0238 & 0.0232 & 0.0238 & 0.0233 & 0.0219 \\
Ours & \textbf{0.0062} & \textbf{0.0074} & 0.0117 & 0.0099 & 0.0080 & \textbf{0.0086} & \textbf{0.0056} & \textbf{0.0049} & \textbf{0.0086} & \textbf{0.0037} & \textbf{0.0043} & \textbf{0.0054} & \textbf{0.0035} & \textbf{0.0067} \\
\hline
\end{tabular}
}
\end{center}

The prediction curves in Figure \ref{fig:[phm2012_rul]} provide visual evidence for the above quantitative results. For most test bearings, the predicted RUL curves follow the global decreasing tendency of the ground-truth RUL, indicating that the proposed model can capture the main degradation evolution pattern across different operating conditions. The curves under Condition II and Condition III show relatively better agreement with the ground truth, which is consistent with the lower C2\_Avg and Br3\_3 errors in Tables~\ref{tab:rmse_phm2012}--\ref{tab:eas_phm2012}. In contrast, several bearings under Condition I exhibit more visible fluctuations and local deviations, which explains why the performance gain under C1\_Avg is smaller than that under Condition II and Condition III. These observations indicate that the bidirectional graph representation, memory-augmented prediction, and KAN-based regressor jointly improve the modeling of long-term degradation trends, while abrupt local variations in some PHM2012 trajectories remain challenging.

\begin{center}
    \includegraphics[width=0.95\textwidth]{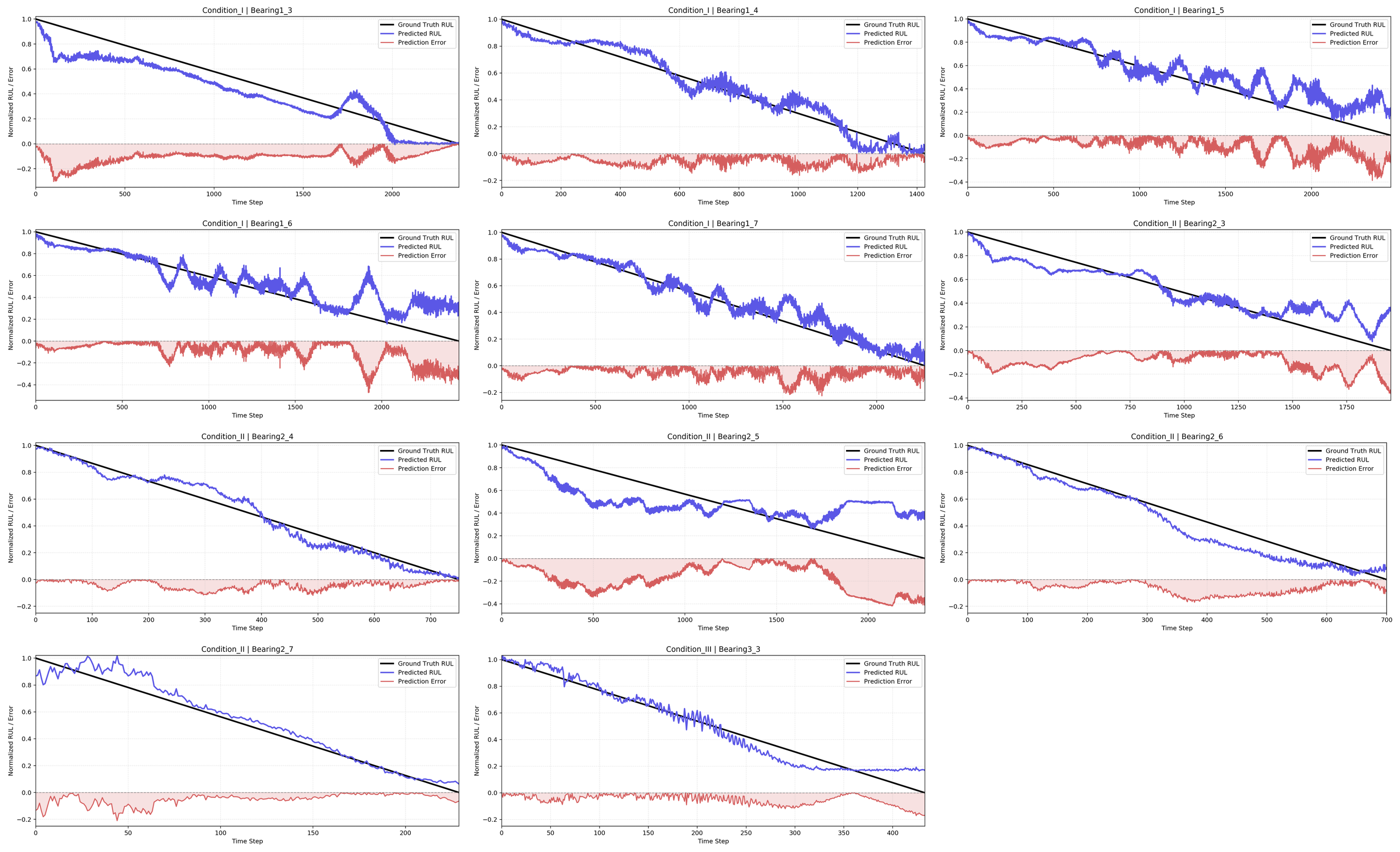}
    \captionof{figure}{RUL prediction results for test bearings in the PHM2012 dataset.}
    \label{fig:[phm2012_rul]}
\end{center}

\subsubsection{Interpretability experiments}

In this section, Condition II is selected as the representative case for interpretability analysis. The functional responses under Condition II are shown in Figure~\ref{fig:phm_condition2_responses}. In the forward graph propagation branch, the low-order functions still preserve clear degradation trend information. The conv1 order 0 mappings exhibit monotonic increasing or smoothly varying patterns, indicating that the forward branch can capture the accumulated degradation tendency along the bearing life cycle. This phenomenon suggests that the degradation evolution under this condition is not a purely smooth process, and the forward propagation needs to model both the global degradation trend and local nonlinear changes. In particular, the higher-order responses show more boundary-sensitive variations, showing multi-order Chebyshev graph propagation is able to identify local sensitive regions and stage-dependent degradation changes.

The backward graph propagation branch presents a complementary functional pattern. In conv1 order 0 and conv2 order 0, increasing and decreasing mappings coexist. This indicates that the backward branch does not simply repeat the forward degradation modeling process, but provides reverse structural constraints from later degradation states to earlier states. Such reverse information is useful for correcting the representation of weak or ambiguous degradation stages. A more evident phenomenon appears in order 1, where both forward and backward branches contain boundary jumps and local peaks. These responses imply that order 1 mainly captures local abrupt changes and boundary-sensitive degradation regions. Compared with order 1, order 2 becomes smoother but still maintains clear nonlinear forms, such as U-shaped, S-shaped, accelerating increasing, and accelerating decreasing curves. Therefore, order 2 acts as a higher-order curvature correction term, which further refines complex local degradation responses rather than merely smoothing the representation.

In contrast, the final KAN regressor produces much smoother functions. This indicates that the final KAN regressor does not directly preserve the local peaks, boundary jumps, and abrupt nonlinear responses learned by the graph branches. Instead, it compresses these complex graph-level functions into lower-complexity and smoother RUL regression mappings. This functional transformation explains why PE-BMGN achieves strong quantitative performance under Condition II: the bidirectional graph propagation branches provide sufficiently rich nonlinear degradation representations, while the final KAN regressor integrates them into stable and interpretable lifetime prediction functions.

\begin{center}
    \includegraphics[width=0.95\textwidth]{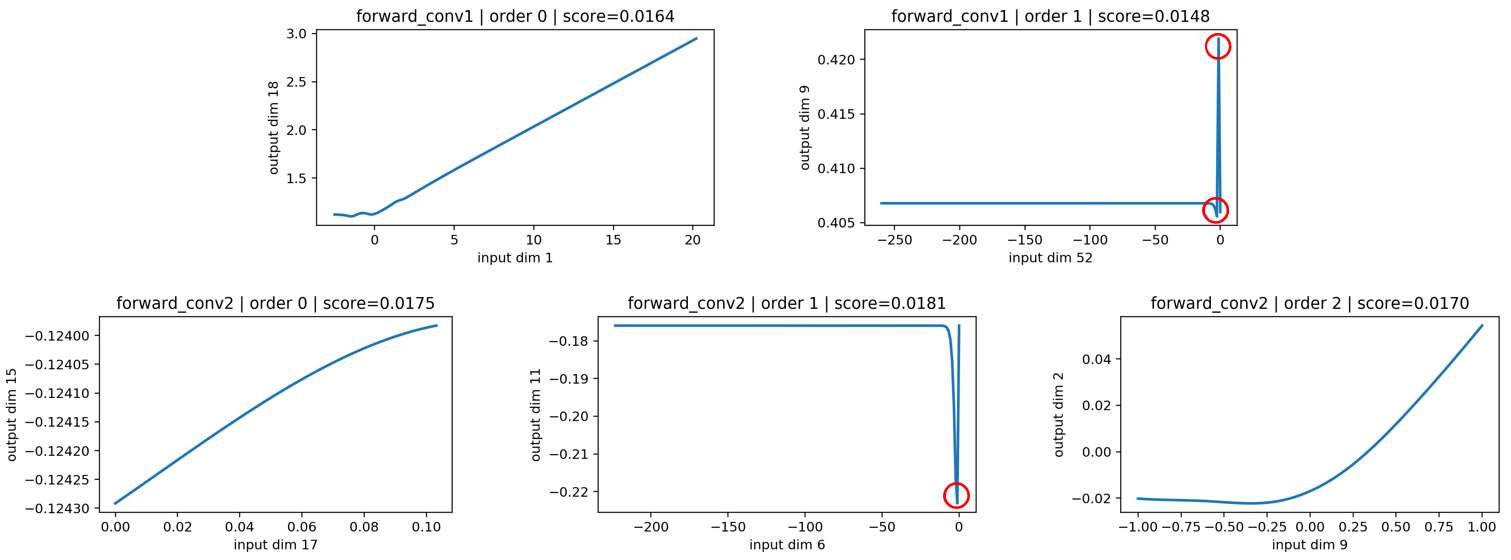}\par
    {\small (a) Forward graph propagation.\par}
    \vspace{0.2em}
    \includegraphics[width=0.95\textwidth]{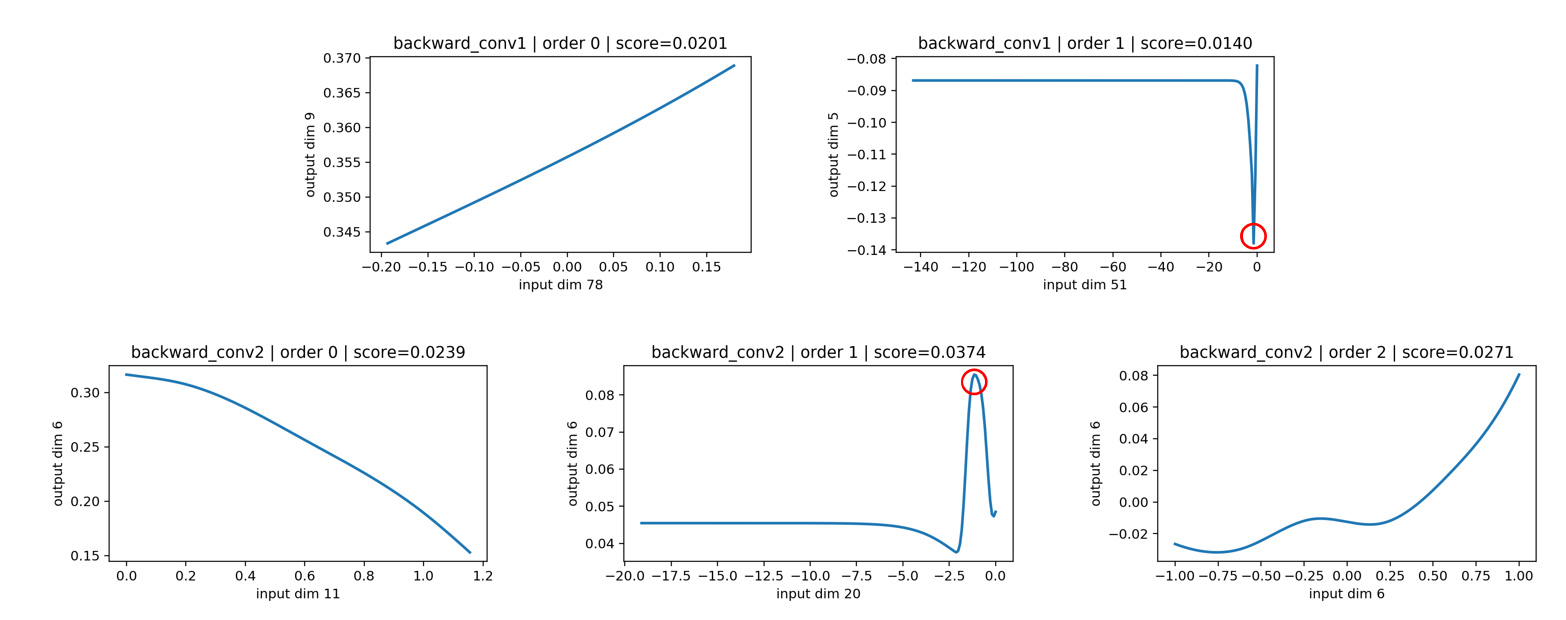}\par
    {\small (b) Backward graph propagation.\par}
    \vspace{0.2em}
    \includegraphics[width=0.7\textwidth]{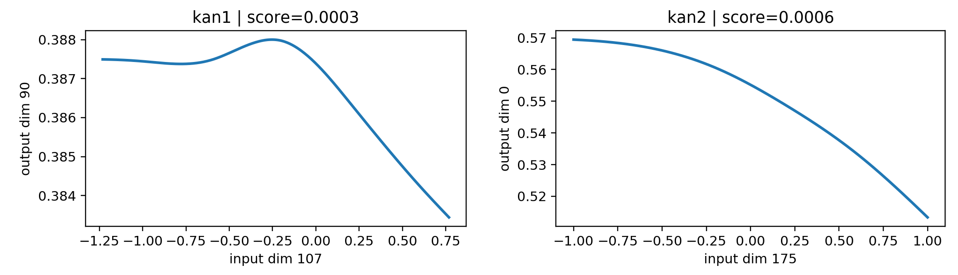}\par
    {\small (c) KAN regressor.\par}
    \captionof{figure}{Learned function responses of Condition II on the PHM2012 dataset.}
    \label{fig:phm_condition2_responses}
\end{center}

\subsection{Ablation experiments}

\subsubsection{Model ablation experiments}

\begin{figure}
    \centering

    \begin{subfigure}{.495\textwidth}
        \centering
        \includegraphics[width=\linewidth,height=.195\textheight,keepaspectratio]{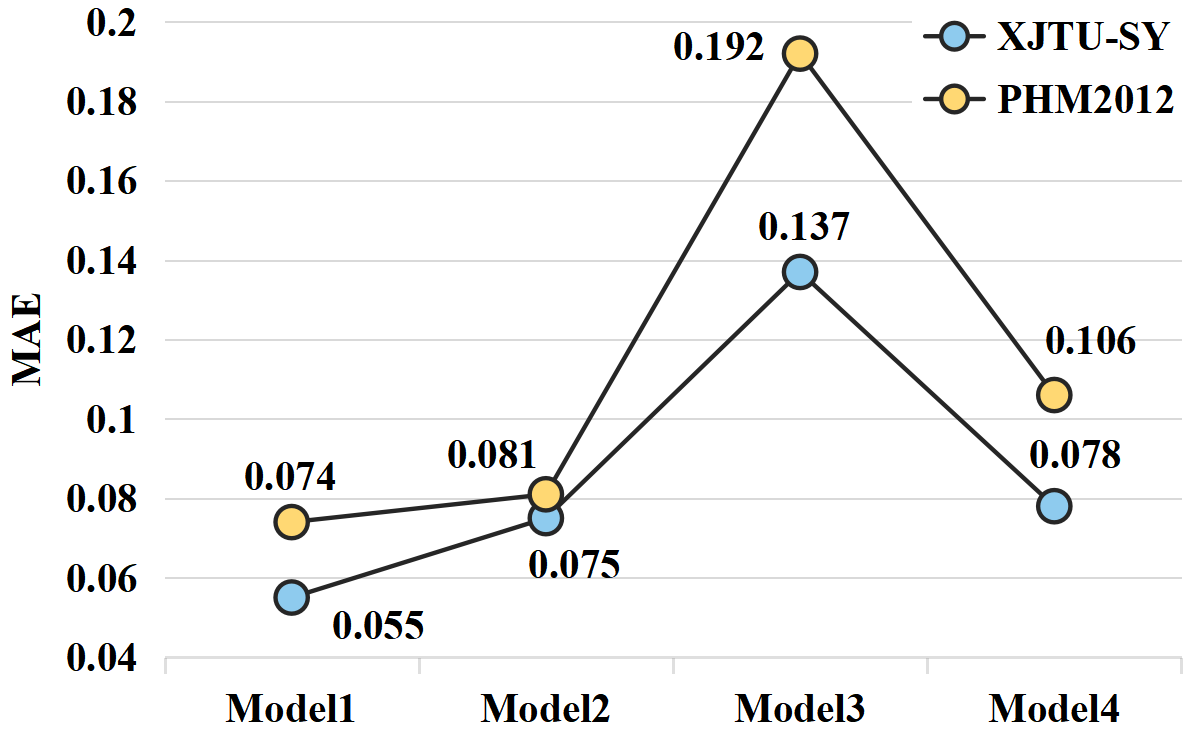}
        \caption{}
        \label{fig:ablation_1a}
    \end{subfigure}%
    \hspace{0.005\textwidth}%
    \begin{subfigure}{.495\textwidth}
        \centering
        \includegraphics[width=\linewidth,height=.195\textheight,keepaspectratio]{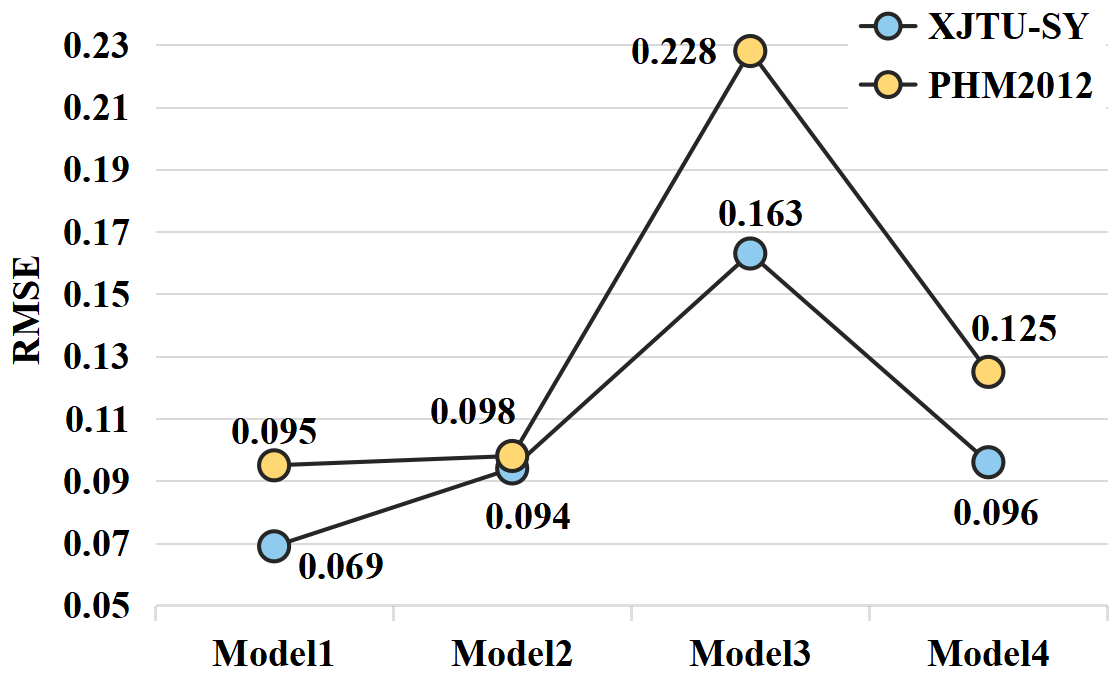}
        \caption{}
        \label{fig:ablation_1b}
    \end{subfigure}

    \vspace{0.15em}

    \begin{subfigure}{.70\textwidth}
        \centering
        \includegraphics[width=\linewidth,height=.185\textheight,keepaspectratio]{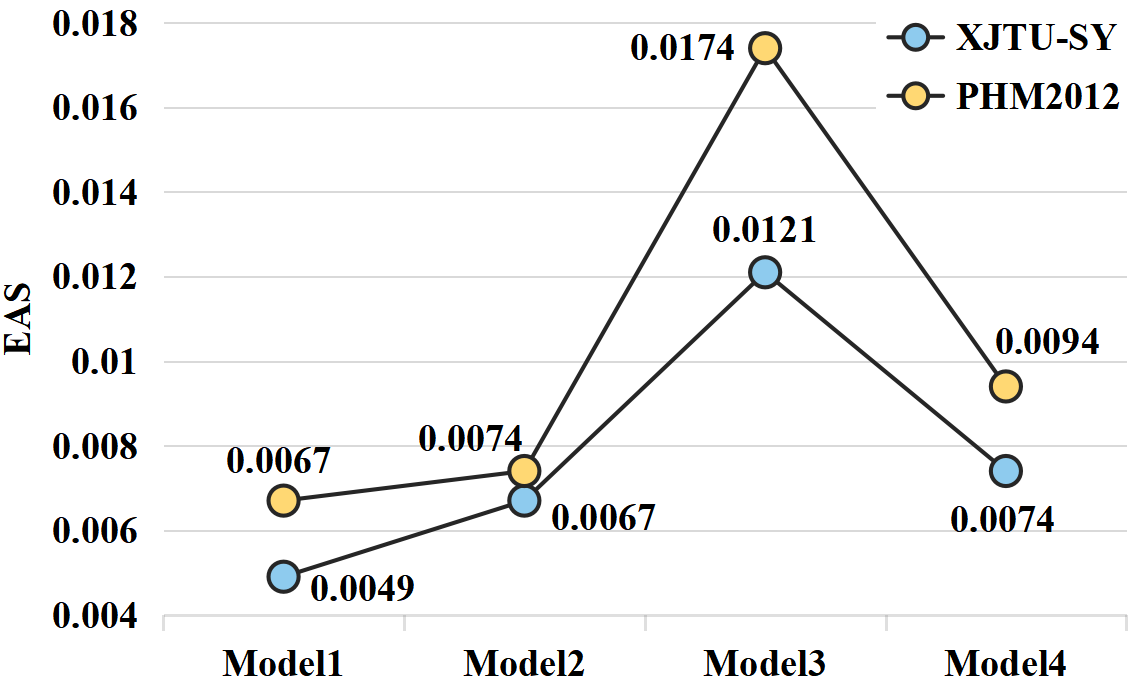}
        \caption{}
        \label{fig:ablation_1c}
    \end{subfigure}

    \caption{Model ablation experimental results. Model1 denotes the complete PE-BMGN, Model2 removes the memory-augmented prediction network, Model3 removes the bidirectional multi-order graph fusion network, and Model4 removes both modules simultaneously.}
    \label{fig:ablation_1}
\end{figure}

To verify the contribution of the main components in PE-BMGN, four model variants are compared in this section. As shown in Figure \ref{fig:ablation_1}, Model1 denotes the complete PE-BMGN, Model2 removes the memory-augmented prediction network, Model3 removes the bidirectional multi-order graph fusion network, and Model4 removes both modules simultaneously. For both datasets and all three metrics, Model1 consistently achieves the lowest prediction errors, indicating that the complete framework provides the most reliable RUL estimation. Specifically, on the XJTU-SY dataset, Model1 obtains RMSE, MAE, and EAS values of 0.069, 0.055, and 0.0049, respectively. On the PHM2012 dataset, the corresponding values are 0.095, 0.074, and 0.0067. These results demonstrate that the bidirectional degradation representation, memory-enhanced prediction, and KAN-based nonlinear regression are complementary rather than isolated components.

After removing the memory-augmented prediction network, Model2 shows higher errors than Model1 on both datasets. On the XJTU-SY dataset, the RMSE, MAE, and EAS increase from 0.069, 0.055, and 0.0049 to 0.094, 0.075, and 0.0067, respectively. On the PHM2012 dataset, the three metrics increase from 0.095, 0.074, and 0.0067 to 0.098, 0.081, and 0.0074, respectively. This indicates that memory retrieval contributes to prediction stability by introducing historical degradation prototypes into the final regression stage. The performance gap is more obvious on the XJTU-SY dataset, suggesting that the memory bank is especially useful when the model needs to distinguish degradation patterns among multiple leave-one-out bearing trajectories. Without this module, the prediction process depends more heavily on the current latent representation, which weakens the ability to exploit historically similar degradation states.

The most significant performance degradation occurs in Model3, where the bidirectional multi-order graph fusion network is removed. Compared with Model1, the RMSE increases from 0.069 to 0.163 on the XJTU-SY dataset and from 0.095 to 0.228 on the PHM2012 dataset. Similar trends can also be observed in MAE and EAS, where Model3 reaches 0.137 and 0.0121 on XJTU-SY, and 0.192 and 0.0174 on PHM2012. This confirms that the bidirectional multi-order graph fusion network is the most critical component for extracting effective degradation dependencies. When this module is removed, the model loses the ability to jointly describe forward and backward information. As a result, the downstream memory retrieval and KAN regression stages cannot obtain sufficiently discriminative degradation representations, leading to much larger numerical errors and engineering risk.

Model4 removes both the bidirectional multi-order graph fusion network and the memory-augmented prediction network. Its performance is worse than Model1 and Model2, but better than Model3 on both datasets. For example, the RMSE values of Model4 are 0.096 and 0.125 on XJTU-SY and PHM2012, respectively, which are lower than those of Model3 but still higher than those of Model1. This phenomenon suggests that the memory module is not universally beneficial when the preceding degradation representation is insufficient. If the bidirectional graph representation is removed, the retrieved memory prototypes may be based on less discriminative features, which can amplify mismatched historical information. Therefore, the memory-augmented prediction network needs to cooperate with the bidirectional multi-order graph fusion network to provide stable performance gains.

\subsubsection{Hyperparameter ablation experiments}

\begin{figure}
    \centering

    \begin{subfigure}{.495\textwidth}
        \centering
        \includegraphics[width=\linewidth,height=.195\textheight,keepaspectratio]{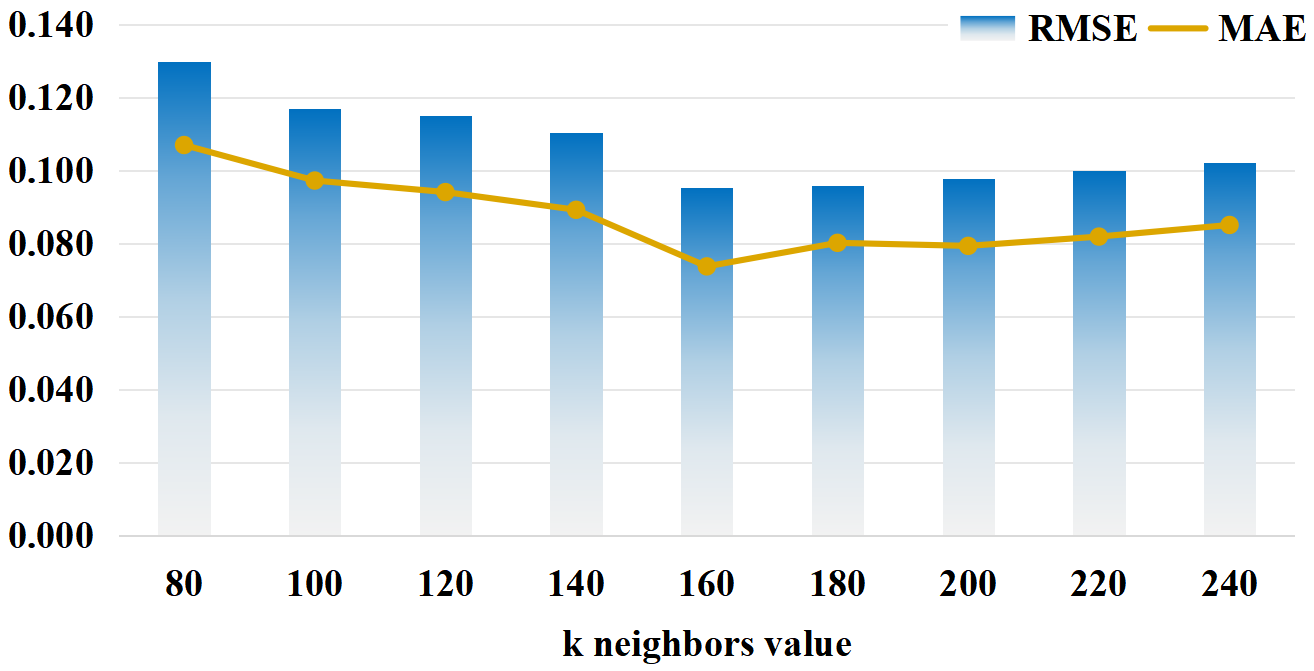}
        \caption{}
        \label{fig:ablation_2a}
    \end{subfigure}%
    \hspace{0.005\textwidth}%
    \begin{subfigure}{.495\textwidth}
        \centering
        \includegraphics[width=\linewidth,height=.195\textheight,keepaspectratio]{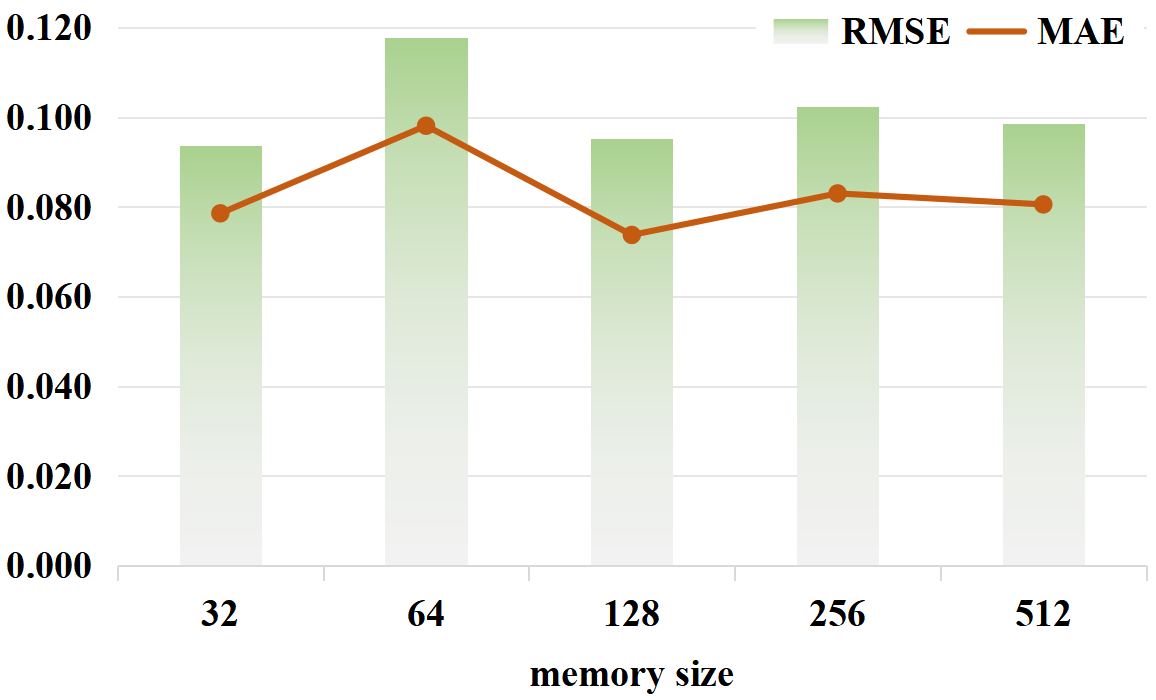}
        \caption{}
        \label{fig:ablation_2b}
    \end{subfigure}

    \caption{The hyperparameter ablation experimental results on the XJTU-SY dataset.}
    \label{fig:ablation_2}
\end{figure}

\begin{figure}
    \centering

    \begin{subfigure}{.495\textwidth}
        \centering
        \includegraphics[width=\linewidth,height=.195\textheight,keepaspectratio]{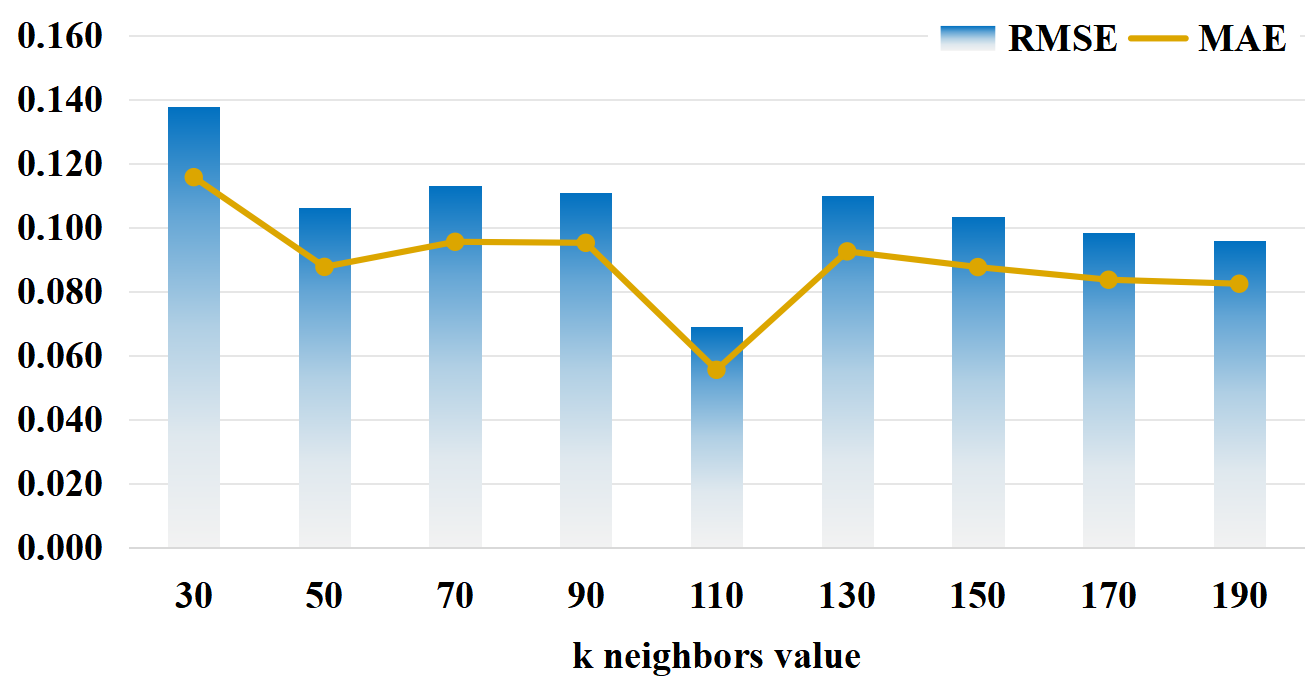}
        \caption{}
        \label{fig:ablation_3a}
    \end{subfigure}%
    \hspace{0.005\textwidth}%
    \begin{subfigure}{.495\textwidth}
        \centering
        \includegraphics[width=\linewidth,height=.195\textheight,keepaspectratio]{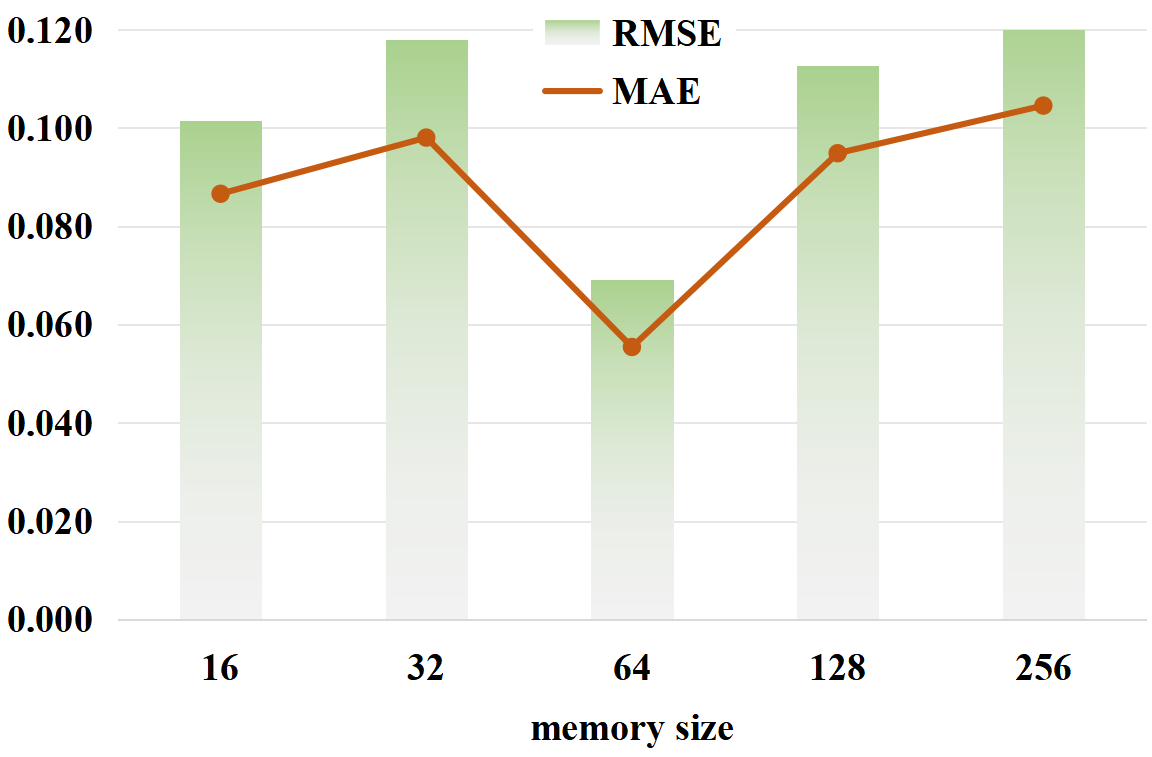}
        \caption{}
        \label{fig:ablation_3b}
    \end{subfigure}

    \caption{The hyperparameter ablation experimental results on the PHM2012 dataset.}
    \label{fig:ablation_3}
\end{figure}

To further investigate the sensitivity of PE-BMGN to key hyperparameters, this section analyzes the influence of the number of retained neighboring nodes $k$ and the memory bank size. The ablation results on the XJTU-SY and PHM2012 datasets are shown in Figure \ref{fig:ablation_2} and Figure \ref{fig:ablation_3}, respectively. These two hyperparameters correspond to two important modeling mechanisms in the proposed framework: $k$ controls the sparsity of the adaptive graph used in the bidirectional multi-order graph fusion network, while the memory bank size determines the capacity of historical degradation prototypes used in the memory-augmented prediction network.

As shown in Figure \ref{fig:ablation_2}(a), the prediction performance on the XJTU-SY dataset is sensitive to the number of retained neighbors. When $k$ increases from 80 to 160, both RMSE and MAE show a clear decreasing trend, indicating that a very sparse graph cannot provide sufficient degradation dependency information for bidirectional graph propagation. When $k$ is set to 160, the model obtains the best overall performance in this group of experiments. However, when $k$ continues to increase from 180 to 240, the errors gradually increase again. This phenomenon suggests that retaining too many neighboring nodes may introduce weakly related or redundant temporal dependencies into graph propagation, which weakens the discriminability of the learned degradation representation. Therefore, an appropriate graph sparsity is necessary: too small $k$ leads to insufficient dependency modeling, while too large $k$ reduces the selectivity of adaptive graph learning.

A similar but dataset-dependent trend can be observed on the PHM2012 dataset in Figure \ref{fig:ablation_3}(a). The model performs poorly when $k$ is too small, such as $k=30$, because the graph structure may fail to cover enough temporal degradation relationships. As $k$ increases, the prediction error generally decreases and reaches the best performance around $k=110$. After that, the model performance becomes worse when $k$ is further enlarged, especially at $k=130$ and $k=150$. This indicates that the PHM2012 dataset also requires a moderate graph density, but its optimal neighbor size is smaller than that of the XJTU-SY dataset. The reason may be that PHM2012 contains shorter signal samples and fewer training bearings under each condition, so an overly dense graph is more likely to introduce noisy or mismatched degradation dependencies. These results confirm that $k$ should be selected according to the temporal scale and degradation complexity of each dataset.

The influence of memory bank size is shown in Figure \ref{fig:ablation_2}(b) and Figure \ref{fig:ablation_3}(b). On the XJTU-SY dataset, the model does not achieve the best performance with the smallest or largest memory capacity. When the memory bank is too small, the stored prototypes may not be sufficient to represent diverse degradation patterns across different bearings. In contrast, when the memory bank becomes too large, the retrieval process may include redundant or weakly matched prototypes, which can reduce the effectiveness of memory-enhanced prediction. The results show that a moderate memory size provides a better balance between prototype diversity and retrieval reliability. In particular, the setting around 128 memory entries provides the most stable overall behavior on the XJTU-SY dataset, with low RMSE and the lowest MAE in this hyperparameter group.

For the PHM2012 dataset, the sensitivity to memory size is more evident. As shown in Figure \ref{fig:ablation_3}(b), the model achieves the lowest RMSE and MAE when the memory bank size is set to 64. Both smaller and larger memory sizes lead to degraded performance. A memory bank with only 16 or 32 entries may not cover enough representative degradation prototypes, while memory sizes of 128 and 256 may introduce excessive candidate prototypes and increase the risk of retrieving less relevant historical patterns. This trend indicates that the memory-augmented prediction network benefits from sufficient but compact prototype storage. An appropriate memory capacity allows the KAN regressor to exploit useful historical degradation knowledge, whereas excessive memory capacity may weaken the precision of prototype retrieval.

\section{Conclusion}

RUL prediction of rolling bearings is an important task for intelligent maintenance. In this paper, a physics-enhanced bidirectional multi-order graph fusion network, PE-BMGN, is proposed for interpretable bearing RUL prediction. PE-BMGN is designed to model the bearing life cycle as an ordered degradation trajectory. The bidirectional multi-order graph fusion network extracts complementary degradation information from forward degradation accumulation and backward structural calibration, while the gated cross fusion mechanism adaptively balances their contributions at different degradation stages. In addition, the memory-augmented prediction network retrieves representative historical degradation prototypes, so that the final prediction is guided not only by the current latent representation but also by reusable degradation knowledge. By introducing Kolmogorov-Arnold functional mappings into the graph transformation and prediction stages, the proposed framework provides a more explicit way to analyze nonlinear degradation-to-RUL mappings. Finally, the physics-enhanced dynamic loss function constrains both physical consistency and functional stability, encouraging the model to learn reliable and compact degradation representations. 

Comprehensive experiments are conducted on the XJTU-SY and PHM2012 bearing datasets. The trajectory-level prediction results further demonstrate that PE-BMGN can capture the global degradation trend for most bearings. Meanwhile, the interpretability analysis shows that the learned functions provide an intuitive view of how forward graph propagation, backward graph propagation, and the final KAN regressor contribute to RUL prediction. Smooth and coherent functional responses are usually associated with accurate predictions, whereas local peaks, boundary-sensitive changes, or inconsistent nonlinear responses reveal the difficulty of atypical degradation trajectories. The ablation results further verify the necessity of the proposed components: the bidirectional multi-order graph fusion network provides the main degradation dependency modeling capability, while the memory-augmented prediction network improves prediction stability by introducing historical degradation knowledge.

In future work, more physically grounded constraints related to bearing fault mechanisms and degradation dynamics will be incorporated into the optimization process. We also plan to use the interpretable analysis to feed back into the design of the model architecture, thereby achieving a more effective RUL prediction method. In addition, adaptive memory updating, uncertainty-aware prototype retrieval, and cross-condition transfer strategies will be further explored to improve the robustness, reliability, and deployability of the proposed framework in real industrial scenarios.

\section*{CRediT authorship contribution statement}
\textbf{Haoxuan Zhang:} Conceptualization; Data curation; Formal analysis; Methodology; Visualization; Writing - original draft. \textbf{Dinghao Yang:} Conceptualization; Data curation; Formal analysis; Methodology; Writing - original draft. \textbf{Kangning Zhang:} Formal analysis; Investigation; Software; Visualization; Writing - review \& editing. \textbf{Shaoyong Guo:} Formal analysis; Investigation; Validation; Writing - review \& editing. \textbf{Haisheng Li:} Funding acquisition; Formal analysis; Investigation; Software; Writing - review \& editing. \textbf{Rui Yang:} Project administration; Resources; Supervision; Validation; Investigation; Writing - review \& editing. \textbf{Ruijun Liu:} Funding acquisition; Resources; Methodology; Supervision; Writing - review \& editing.

\section*{Acknowledgments}
This work is supported by the National Natural Science Foundation of China (No. U25A20465, No. U25A20446) and Open Foundation of State key Laboratory of Networking and Switching Technology (Beijing University of Posts and Telecommunications) (No. SKLNST-2021-1-09).

\bibliographystyle{unsrtnat}
\bibliography{cas-refs}

\end{document}